\documentclass[twocolumn]{aastex631}

\usepackage{amsmath}
\usepackage{makecell}

\usepackage{scalerel}

\newcommand{\RN}[1]{\textup{\uppercase\expandafter{\romannumeral#1}}}

\def\NHUNIT{\ifmmode {\rm \,cm^{-2}} \else $\rm \,cm^{-2}$ \fi}

\usepackage{xcolor}
\usepackage{diagbox}
\usepackage{txfonts}
\newcommand{\blos}{$B_{\rm{LOS}}$}
\newcommand{\bpos}{$B_{\rm{POS}}$}
\newcommand{\rmref}{RM$_{\rm{ref}}$}
\newcommand{\rmon}{RM$_{\rm{ON}}$}
\newcommand{\rmoff}{RM$_{\rm{OFF}}$}
\newcommand{\rmmc}{RM$_{\rm{MC}}$}

\newcommand{\Av}{$A_{{V}}$}
\newcommand{\Avth}{A$_{\rm{V,th}}$}
\newcommand{\Avon}{A$_{\rm{V,ON}}$}

\newcommand{\Avref}{A$_{\rm{V,ref}}$}
\usepackage{scalerel}

\newcommand{\Ne}{$N_{e}$}

\usepackage{tikz}
\usetikzlibrary{shapes.geometric, arrows}
\tikzstyle{startstop} = [rectangle, rounded corners, minimum width=3cm, minimum height=1cm,text centered, draw=black, fill=white!30]
\tikzstyle{io} = [trapezium, trapezium left angle=70, trapezium right angle=110, minimum width=3cm, minimum height=1cm, text centered, draw=black, fill=white!30]
\tikzstyle{process} = [rectangle, minimum width=3cm, minimum height=1cm, text centered, draw=black, fill=white!30]
\tikzstyle{invisible} = [rectangle, minimum width=3cm, minimum height=0.1cm, text centered, draw=white, fill=white!30]
\tikzstyle{decision} = [diamond, minimum width=3cm, minimum height=0.5cm, text centered, draw=black, fill=white!30]
\tikzstyle{begin} = [draw, ellipse, minimum width=3cm, minimum height=02cm]
\tikzstyle{arrow} = [thick,->,>=stealth]
\usepackage[none]{hyphenat}

\received{}
\revised{}
\accepted{}
\submitjournal{ApJ}

\shorttitle{MC-BLOS}
\shortauthors{Tahani et al.}
\graphicspath{{./}{figures/}}

\begin{document}

\title{MC-BLOS\footnote{software currently available at \url{https://github.com/MehrnooshTahani/MC-BLOS}}: Determination of the Line-of-Sight Component of Magnetic Fields Associated with Molecular Clouds}

\author[0000-0001-8749-1436]{Mehrnoosh Tahani}

\affiliation{Banting Fellow: Kavli Institute for Particle Astrophysics \& Cosmology (KIPAC), Stanford University, Stanford, CA 94305, USA}
\affiliation{Department of Physics, Stanford University, Stanford, CA 94305, USA}
\affiliation{Dominion Radio Astrophysical Observatory, Herzberg Astronomy and Astrophysics Research Centre, National Research Council Canada, P. O. Box 248, Penticton, BC V2A 6J9 Canada}
\author{John Ming Ngo}
\affiliation{Department of Physics, University of Alberta, Edmonton, Alberta T6G 2E1, Canada}
\affiliation{Dominion Radio Astrophysical Observatory, Herzberg Astronomy and Astrophysics Research Centre, National Research Council Canada, P. O. Box 248, Penticton, BC V2A 6J9 Canada}
\author[0009-0001-3879-3910]{Jennifer Glover}
\affiliation{Department of Physics, McGill University, Montr\'eal,  Quebec H3A 2T8, Canada}
\affiliation{Dominion Radio Astrophysical Observatory, Herzberg Astronomy and Astrophysics Research Centre, National Research Council Canada, P. O. Box 248, Penticton, BC V2A 6J9 Canada}
\author{Ryan Clairmont}
\affiliation{Department of Physics, Stanford University, Stanford, CA 94305, USA}
\author{Gabriel M. Zarazua}
\affiliation{Department of Physics and Astronomy, San Francisco State University, 1600 Holloway Ave., San Francisco, CA 94132, USA}
\author{René Plume}
\affiliation{Department of Physics and Astronomy, University of Calgary, 2500 University Drive NW, Calgary, Alberta T2N 1N4, Canada}

\begin{abstract}

In recent years a number of surveys and telescopes have observed the plane-of-sky component of magnetic fields associated with molecular clouds. However, observations of their line-of-sight magnetic field remain limited. To address this issue, \citet{Tahanietal2018} developed a technique based on Faraday rotation. The technique incorporates an ON-OFF approach to identify the rotation measure induced by the magnetic fields associated with the cloud. The upcoming abundance of Faraday rotation observations from the Square Kilometer Array and its pathfinders necessitates robustly-tested software to automatically obtain line-of-sight magnetic fields of molecular clouds.
We developed software, called MC-BLOS (Molecular Cloud Line-of-Sight Magnetic Field), to carry out the technique in an automated manner. The software's input are Faraday rotation of point sources (extra-galactic sources or pulsars), extinction or column density maps, chemical evolution code results, and a text/CSV file, which allows the user to specify the cloud name or other parameters pertaining to the technique.
For each cloud, the software invokes a set of predefined initial parameters such as density, temperature, and surrounding boundary, which the user can modify. The software then runs the technique automatically, outputting line-of-sight magnetic field maps and tables (including uncertainties) at the end of the process. This automated approach significantly reduces analysis time compared to manual methods. We have tested the software on previously-published clouds, and the results are consistent within the reported uncertainty range. This software will facilitate the analysis of forthcoming Faraday rotation observations, enabling a better understanding of the role of magnetic fields in molecular cloud dynamics and star formation.
\end{abstract}

   
%
\section{Introduction}
\label{sec:introduction}

Mapping the three-dimensional (3D) structure of magnetic fields is crucial for understanding the role of magnetic fields in the formation and evolution of molecular clouds and their subsequent star formation. While magnetic fields can hinder star formation by providing additional support against gravitational collapse~\citep{SeifriedWalch2015} or when combined with feedback mechanism such as stellar outflows~\citep{KrumholzFederrath2019}, they are also necessary for regulating core collapse through magnetic braking~\citep{PudritzRay2019}. Recent studies have shown that the relative orientation between magnetic fields and density structures such as molecular clouds can provide insights into their formation and evolution~\citep{PlanckXXXV, Arzoumanianetal2021, Pattleetal2023PP7}. However, to fully understand the role of magnetic fields in molecular cloud dynamics and star formation, it is essential to observe and analyze their 3D structure, including both the plane-of-sky (\bpos ) and line-of-sight (\blos ) magnetic field components~\citep{Tahanietal2019, Tahanietal2022O, Tahanietal2022P, Tahani2022}.

Far-infrared (FIR) and sub-millimeter observations, such as those by Planck~\citep[e.g.,][]{PlanckXIX2015} or James Clerk Maxwell telescope~\citep[e.g.,][]{Arzoumanianetal2021, Hwangetal2022, Chingetal2022, Tahanietal2023}, have successfully used dust emission polarization in large surveys to map \bpos  .  In this technique  the measured polarization is perpendicular to the magnetic field in the plane of the sky~\citep{PattleFissel2019}, and \bpos\   is averaged along the lines sight~\citep[mass weighted;][]{Seifriedetal2019}. Observations using dust extinction polarization at near-infrared (NIR) and optical wavelengths, referred to as the starlight polarization technique, combined with stellar distances obtained by Gaia, enable us to obtain a tomography of the two-dimensional magnetic field along the line of sight ~\citep[tomographic \bpos ;][]{Doietal2021, Panopoulouetal2023, Doietal2024}. This technique involves measuring the polarization of starlight that has passed through the interstellar medium, where the interstellar dust grain have aligned their long axis perpendicular to magnetic field lines~\citep{LazarianHoang2007}. In this case the measured polarized light is parallel to the plane-of-sky magnetic field orientation. By combining these measurements with stellar distances, it is possible to construct a tomographic view of the \bpos\ orientation along the line of sight. However, to study the true 3D magnetic fields, observing \blos\  is crucial.  

Zeeman splitting is sensitive to the line-of-sight component of magnetic fields and is the most direct method for measuring the strength of magnetic fields~\citep{CrutcherKemball2019}, but it requires long telescope integration times and  strong magnetic fields~\citep{Crutcher2012, RobishawThesis}. Faraday rotation is a more widely applicable technique for determining \blos , as it does not have the same limitations as Zeeman splitting. However, it is less direct (more model dependent) for inferring the field strengths.

Faraday rotation is a powerful technique for determining \blos\ and refers to the rotation of the polarization plane of a linearly polarized electromagnetic wave in a magneto-ionic region. The amount of rotation in the medium is described by
\begin{equation}
\label{FaradayRot}
\Delta \Psi = \lambda^2 \bigg(0.812 \int n_{\text{e}} \mathbf{B \cdot dl} \bigg) = \lambda ^2 \rm \text{RM} \rm~[rad],
\end{equation}
where $\Delta \Psi$ [rad], $\lambda$ [m], $\mathbf{B}$ [$\mu$G], $\mathbf{dl}$ [pc], and $n_{\text{e}}$ [cm$^{-3}$] are the amount of rotation in polarization angle, wavelength of the electromagnetic wave, external magnetic field, path length through the magnetized region, and electron number density of the region, respectively. The quantity in brackets is an integral of the product of the electron density and the magnetic field along the path length, and it is known as the rotation measure (RM).

The average \blos\ of a cloud or region can be determined by observing its induced RM, where the RM's sign indicates the \blos\ direction. This is achieved by studying the observed polarization angle with respect to $\lambda^2$ over a range of wavelengths~\citep[e.g.,][]{BrownTaylor2001}. Assuming \blos\ is constant within the region and can be taken out of the integral, only the electron column density of the region needs to be determined to find \blos :
\begin{equation}
B_{\rm{LOS}} = \frac{RM}{\int n_e dl} = \frac{RM}{N_e},
\label{eq:findBlos}
\end{equation}
where $N_e$ is the electron column density of the region.

However, until relatively recently, despite previous attempts~\citep{WollebenReich2004Molecular}, Faraday rotation had not been successfully applied to study \blos\ in molecular clouds.  
\citet{Tahanietal2018} demonstrated the feasibility of using Faraday rotation to determine \blos\ associated with molecular clouds. Their technique (hereafter MC-BLOS method) uses the RM value of  point sources (such as unresolved pulsars, or radio galaxies) and an ON-OFF approach to find the RM induced by the cloud and effectively separate the RM contribution of the magnetized molecular cloud from that of the foreground and background regions. This enables the determination of  \blos\ direction associated with the molecular cloud at each point. To determine the electron densities and hence the strength of magnetic fields, the technique uses a chemical evolution code and extinction (\Av\ or column density) maps to model the ionization state of the cloud. 

The MC-BLOS technique has been applied to study and map the \blos\ of several molecular clouds, and the results have been found to be consistent with available Zeeman measurements~\citep{Tahanietal2018}. These \blos\ maps have since been applied to study and reconstruct the 3D magnetic field structure in the Orion A and Perseus molecular clouds~\citep{Tahanietal2022P, Tahanietal2022O}. These studies are the first to provide a complete picture of the 3D magnetic field vectors in molecular clouds, including the \blos\ direction and the \bpos\ orientation and direction (i.e., complete 3D). 

Using the 3D magnetic field structure of the Perseus molecular cloud, \citet{Tahanietal2022P} proposed the cloud's interaction with a previously-unidentified structure, presence of which was later confirmed by \citet{Kounkeletal2022} using kinematic observations. These studies demonstrate the power of the MC-BLOS technique in revealing 3D magnetic fields and hence new insights into the complex interplay between magnetic fields and cloud dynamics.

New and upcoming RM observations, such as those from POSSUM~\citep[][]{Gaensler2010, Vanderwoudeetal2024}, LOFAR~\citep{VanEcketal2017, O'Sullivanetal2023}, and Square Kilometer Array~\citep[SKA;][]{Healdetal2020}, necessitate a software and upgraded technique to determine the \blos\ associated with molecular clouds in an automated manner. 

In this paper, we introduce the python-based, open-source software, MC-BLOS (v1.0), that performs the MC-BLOS technique automatically. 
Additionally, we have made upgrades to the technique to include improvements for better sampling of non-cloud contributions to the RM, such as an optimized algorithm for selecting final OFF positions (referred to as reference points\footnote{we refer to  the the final selected OFF points as reference points. Reference and ``selected OFF" points may be used interchangeably.}) and a more robust method for estimating the background RM. Furthermore, the new software is capable of performing the technique on multiple clouds in one run,  enhancing its efficiency and usability. These upgrades make MC-BLOS (v1.0) a valuable tool for analyzing the upcoming wealth of RM data and studying the 3D magnetic field structure of molecular clouds. We discuss the MC-BLOS technique, the software (including the upgrades), the results, and a summary in sections~\ref{sec:theory}, \ref{sec:software}, \ref{sec:Results}, and \ref{Sec:summary}, respectively.

\section{General theory}
\label{sec:theory}

The MC-BLOS technique consists of two main components: direction determination and strength determination. The direction determination component isolates the RM contribution of the molecular cloud from that of the foreground and background regions (Galactic contribution), while the strength determination component estimates the magnetic field strength using electron density information derived from chemical evolution models and extinction maps. This separation allows the MC-BLOS technique to effectively constrain both the direction and strength of the line-of-sight magnetic field in molecular clouds.

\subsection{Determining the RM induced by the cloud}
\label{sec:findingRMref}

The RM induced by the cloud can be estimated by averaging the RM values of reference points, which are located near the cloud but far enough to be unaffected by its magnetic fields. OFF points are selected from regions with low column densities (e.g., $A_V < 1$\,mag) that are not near high-column density regions. Averaging OFF points from different areas around the cloud ensures proper sampling of the Galactic contribution and that ON and OFF points probe similar path lengths.

\citet{Tahanietal2018} define the average of reference positions as the reference RM (\rmref ). They first identify all the potential OFF positions (N points) by taking a threshold value for \Av\ and finding the RM points that have an associated \Av\ of less than the threshold. 
To find the optimal number of OFF positions (i.e., the number of points needed until the \blos\ values stabilize), they introduce a ``stability trend" analysis, studying how derived magnetic field strength and direction change as the number of OFF positions increases from 1 to N .

Initially, with few OFF positions, there is large variance in the results. As the number increases, variations decrease, and magnetic field strengths and directions stabilize. 
This approach ensures an accurate estimation of the Galactic RM contribution using sufficient OFF positions, while retaining enough RMs for \blos\ determination. 
In this work, we refer to the RM of each individual OFF and ON point as \rmoff\ and \rmon , respectively, while \rmref\ represents the final estimated Galactic contribution to the RM.

In the upgraded version (see Section~\ref{sec:software}), we improve upon the stability trend by automating the selection of reference positions and by using a more robust algorithm to identify the optimal number of OFF positions (see Section~\ref{sec:DirDetermin} for details). Furthermore, we add additional steps to improve the selection of reference points. Subtracting this \rmref\ from the \rmon\ at each point gives us the RM contribution associated with the cloud's magnetic field at each ON position.

\subsection{Determining the electron column density}
\label{sec:findingNe}

After determining the RM contribution of the cloud at each position, the electron column densities are needed to estimate the magnetic field strengths. To this end a chemical evolution code and extinction (column density) maps are required.  \citet{Tahanietal2018} used a chemical evolution code previously tested and utilized by \citet{Gibsonetal2009}.

The chemical evolution code used in the MC-BLOS technique takes into account various input parameters, such as gas density, temperature, UV field strength, and cosmic ray ionization rate. These parameters affect the electron abundance calculations and need to be carefully constrained for each molecular cloud. The code uses the UMIST Rate 99 database for reaction rates \citep{LeTeuffetal1999} and assumes a constant density, temperature, UV field, and cosmic ray ionization rate for each cloud, which can be set as input parameters. The cloud is taken to be homogeneous and planar, divided into 100 layers of equal width. 
The electron column density for a given position cannot be obtained from the electron abundance of a single layer multiplied by the hydrogen column density, as the electron abundance may vary  in each layer.

\begin{figure}[htbp]
 \centering
 \includegraphics[scale=0.3, trim={0cm 0cm 0cm 0cm},clip]{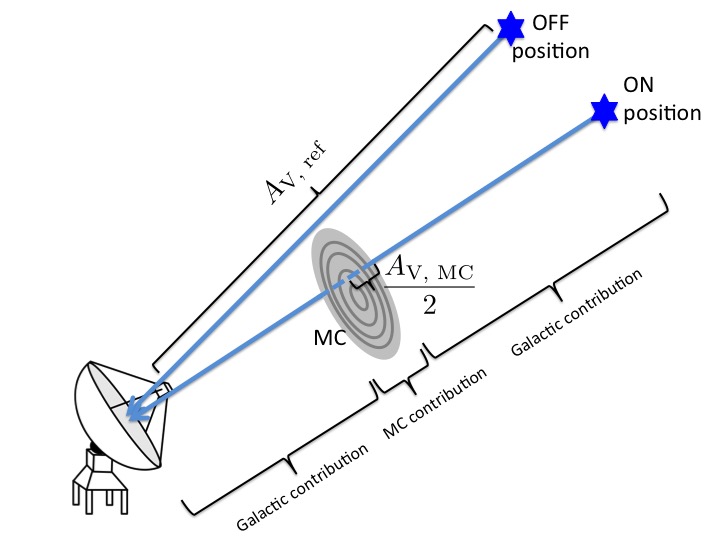}
 \caption{Cloud division into layers for accurate electron column density estimation. Assuming a symmetrical cloud, the electron column density is calculated for each layer up to the center of the cloud along the line of sight, reaching $\frac{A_{V, \text{MC}}}{2}$.} 
 \label{fig:cloudLayerCartoon}
 \end{figure}

MC-BLOS first determines the extinction of each ON position and then calculates the electron column density of all cloud layers up to the point where half the extinction is reached, assuming cloud symmetry along the line of sight, as shown in Figure~\ref{fig:cloudLayerCartoon}. The total electron column density of the ON position is obtained by summing the electron column densities of all layers and multiplying by two, as follows:
\begin{equation}
\begin{aligned}
N_e = & 2 \times \Sigma N_{e, i} \\= & 2 \times \Sigma^{\frac{A_{\text{V, MC}}}{2}} \big((A_{V, i} - A_{V, i - 1}) \times X_{e, i} \big) \times \text{\Av\ to N conversion},
\label{eq:FinalNe}
\end{aligned}
\end{equation}
where $N_{e, i}$ and $X_{e, i}$ are the electron column density and abundance at layer $i$, respectively. $A_{V, i}$ is the extinction at the same layer.

Finally, it determines \blos\ using the total electron column density and \rmref. The upgraded version of the software (Section~\ref{sec:strengthDetermin}) allows users to input their own chemical evolution code and provides default input parameters for many commonly studied molecular clouds, such as those observed by Herschel\footnote{\url{http://www.herschel.fr/cea/gouldbelt/en/Phocea/Vie_des_labos/Ast/ast_visu.php?id_ast=66}}.

\subsection{Determining Uncertainties}
\label{sec:uncertainty}

To estimate the total uncertainties, \cite{Tahanietal2018} consider the following contributions: 
a) \rmon\ uncertainty and standard deviation of \rmref\ ($\frac{\delta RM}{RM}$), 
b) electron density estimation using the chemical evolution code input values, particularly cloud-specific ones such as temperature ($\frac{\delta T}{T}$) and density ($\frac{\delta n({\text{HI}+\text{H}_2})}{n({\text{HI}+\text{H}_2})}$), and 
c) uncertainties due to overlapping RM values on extinction (or column density) maps ($\frac{\delta A_V}{A_V}$). 

They estimate uncertainties in the chemical evolution code input values by varying each parameter within a reasonable range (e.g., $\pm20\%$ for temperature, $\pm50\%$ for density) and calculating the corresponding change in magnetic field. Uncertainties due to overlapping RM values on extinction maps are estimated by comparing extinction values of nearby pixels and propagating uncertainty caused by extinction differences through the magnetic field strength calculation. The total uncertainty is given by:
\begin{equation}
\begin{aligned}
\delta B_{\text{LOS}} = B_{\text{LOS}} \bigg((\frac{\delta RM}{RM})^2  + (\frac{\delta A_V}{A_V})^2&& \\+ (\frac{\delta n({\text{HI}+\text{H}_2})}{n({\text{HI}+\text{H}_2})})^2 + (\frac{\delta T}{T})^2 \bigg)^{1/2}&&. 
\label{beginUncertainty}
\end{aligned}
\end{equation}

To find the \blos\ uncertainty from RM at each point, we use:
\begin{equation}
\begin{aligned}
\Delta B_{RM} = B_{\text{LOS}} \bigg(\frac{\delta (RM_{\text{ref}}) + \delta (RM_{ON})}{RM_{ON} - RM_{\text{ref}}}\bigg),
\end{aligned}
\end{equation}
where $\Delta B_{RM}$ is the uncertainty in \blos\ caused by RM (at each point and associated with \rmref ), $\delta(RM_{\text{ref}})$ is the standard error (standard deviation/$\sqrt N$) of the selected OFF positions, and $\delta(RM_{ON})$ is the tabulated uncertainty of the RM at each ON point (from the RM catalog).

The presented software follows the same uncertainty determination approach as described in \citet{Tahanietal2018} with new upgrades discussed in Section~\ref{sec:uncertaintyUpgrade}. These upgrades include: 1) ensuring that while the uncertainties caused by the RM or the \rmref\ are allowed to result in a change of direction, the uncertainties due to the chemical code or extinction can only influence the strength of \blos\ and not its sign; 2) ensuring that the ON points have an \Av\ higher than $3\times$\Avref.

\section{Software}
\label{sec:software}

\subsection{Data files} 
\label{sec:dataFiles}

To run the software, column density or extinction maps of molecular clouds and an RM catalog are needed. Upon initializing the software (running Initialize.py), cloud maps and the \citet{Tayloretal2009}\footnote{\url{https://cdsarc.cds.unistra.fr/ftp/J/ApJ/702/1230/}} RM catalog are downloaded, as well as the \citet{VanEcketal2023} consolidated RM catalog\footnote{CIRADA-tools; \url{https://github.com/CIRADA-Tools/RMTable}}. Any RM catalog with a format consistent with \citet{VanEcketal2023} or \citet{Tayloretal2009} can be used in the software. The cloud maps are mainly downloaded from publicly-available Herschel observations such as the Herschel Gould Belt Survey\footnote{\url{http://www.herschel.fr/cea/gouldbelt/en/Phocea/Vie_des_labos/Ast/ast_visu.php?id_ast=66}}. 
Users can input their own cloud map (either extinction or column density) or rotation measure catalogs by saving them in the Data directory and specifying the file names in the relevant config files.

\subsection{\blos\ Direction determination}
\label{sec:DirDetermin}

This section describes how the software determines \rmref , which sets the direction of the magnetic field at each ON point associated with the cloud. 
The rotation measure induced by the cloud at any ON position is determined by:
\begin{equation}
RM_{MC} = RM_{ON} - RM_{ref}.  
\label{eq:RMMC}
\end{equation}
Therefore, if \rmmc\ is positive (negative), the direction of the magnetic field associated with the cloud at that position is pointing toward (away from) us, following the convention used in the Faraday rotation measure community. After finding \rmmc\ for each ON point, we can also find the strength of the magnetic field at that point (see Section~\ref{sec:strengthDetermin}).

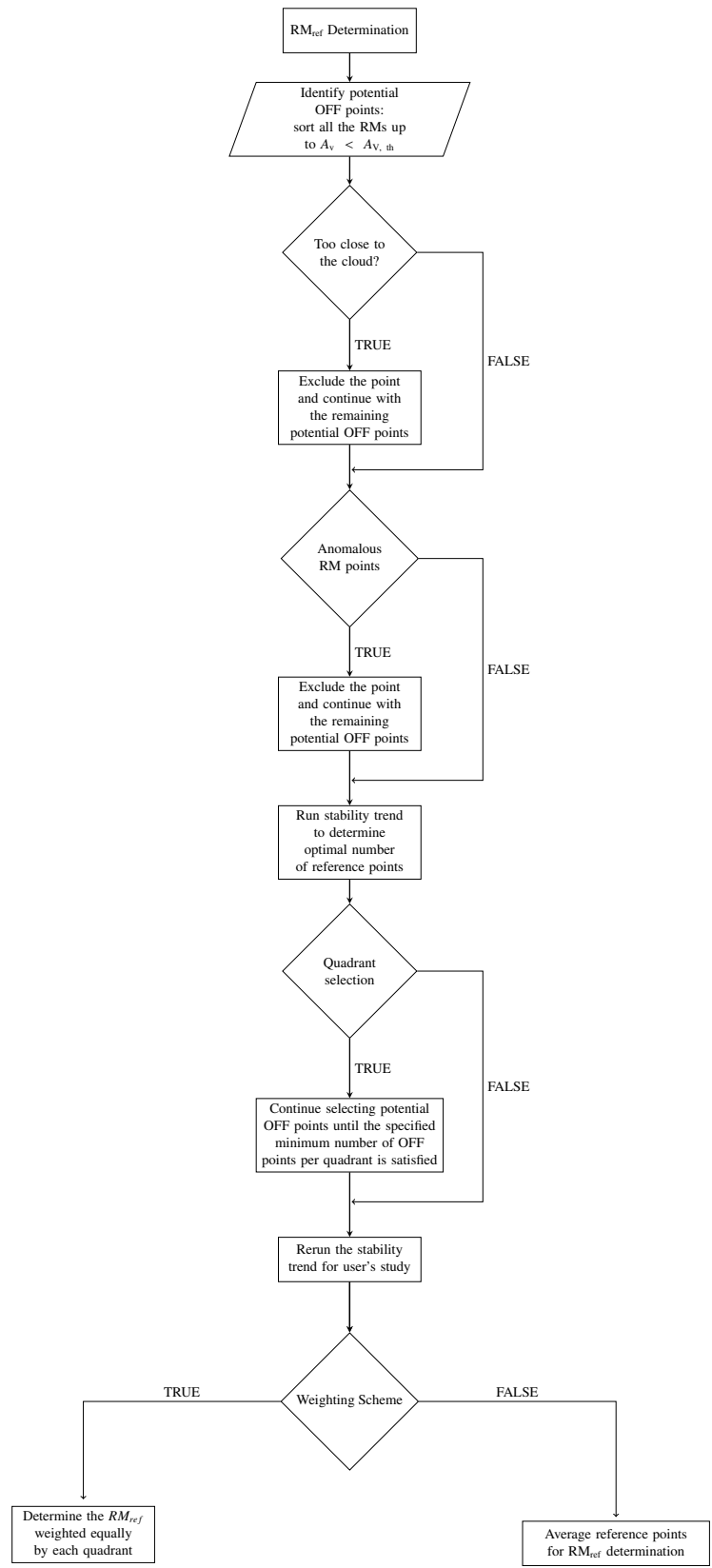
\begin{figure*}
    \centering
    \scalebox{0.62}{
        \begin{tikzpicture}[node distance=2cm]
    \node (init) [invisible] {};
    \node (start) [process, below of=init, xshift = 10cm] {RM$_{\rm{ref}}$ Determination};
    
    \node (sort) [io, below of=start, text width=4cm] {Identify potential \\ OFF points: \\sort all the RMs up \\to $A_{\rm{v}} < A_{\rm{V,~th}}$};
    
    \node (close_to_cloud) [decision, below of=sort, text width=2cm, yshift=-1cm] {Too close to the cloud?};
    
    \node (exclude_close1) [process, below of =close_to_cloud, yshift=-1.5cm, text width = 3cm] {Exclude the point and continue with the remaining potential OFF points};
    
    \node (anomalous) [decision, below of=exclude_close1, yshift=-1.4cm, text width = 2cm] {Anomalous RM points};

     \node (exclude_close2) [process, below of =anomalous, yshift=-1.5cm, text width = 3cm] {Exclude the point and continue with the remaining potential OFF points};
    
    \node (stability_trend) [process, below of=exclude_close2, yshift=-0.9cm, text width = 3 cm] {Run stability trend to determine optimal number of reference points};
    
    \node (quadrant_selection) [decision, below of=stability_trend, yshift=-0.9cm, text width = 2cm] {Quadrant selection};
    
    \node (continue_selecting) [process, below of=quadrant_selection, text width=4cm, yshift=-1.7cm] {Continue selecting potential OFF points until the specified minimum number of OFF points per quadrant is satisfied};
    
    \node (rerun_stability) [process, below of=continue_selecting, yshift=-0.8cm, text width = 3cm] {Rerun the stability trend for user's study};
    
    \node (weighting_scheme) [decision, below of=rerun_stability, yshift=-1.2cm] {Weighting Scheme};
    
    \node (weighted_average) [process, below of=weighting_scheme, xshift=6cm, yshift = -1.2cm, text width=4cm] {Average reference points for \rmref\ determination};
    
    \node (average) [process, below of=weighting_scheme, xshift=-6cm, text width = 3cm, yshift = -1cm] {Determine the $RM_{ref}$ \\ weighted equally by each quadrant};

     \draw [arrow] (start) -- (sort);
     \draw [arrow] (sort) -- (close_to_cloud);
     \draw [arrow] (close_to_cloud) -- node[anchor=west] {TRUE} (exclude_close1);
     \draw [arrow] (exclude_close1) -- (anomalous);
     
     \draw [->](close_to_cloud) -- ++(3,0) --node[anchor=west] {FALSE} ++(0,-4.9) -- ++(-2.95,0);
     
     \draw [arrow] (anomalous) -- node[anchor=west] {TRUE} (exclude_close2);

     \draw [->](anomalous) -- ++(3,0) --node[anchor=west] {FALSE} ++(0,-5) -- ++(-2.95,0);

     \draw [arrow] (exclude_close2) -- (stability_trend);
     \draw [arrow] (stability_trend) -- (quadrant_selection);
     \draw [arrow] (quadrant_selection) -- node[anchor=west] {TRUE} (continue_selecting);
     \draw [->](quadrant_selection) -- ++(3,0) --node[anchor=west] {FALSE} ++(0,-5.2) -- ++(-2.95,0);
     \draw [arrow] (continue_selecting) -- (rerun_stability);
     \draw [arrow] (rerun_stability) -- (weighting_scheme);
     \draw [arrow] (rerun_stability) --  (weighting_scheme);
     \draw [->](weighting_scheme) --node[anchor=south] {FALSE} ++(6,0) -- ++(0,-2.6);
     \draw [->](weighting_scheme) --node[anchor=south] {TRUE} ++(-6,0) -- ++(0,-2.2);
 \end{tikzpicture}
    }
    \caption{Flowchart showing the direction determination steps. The software finds \rmref\ by identifying potential OFF points, excluding anomalous ones, finding the optimal number of OFF points, and finally averaging them.}
\label{flow:DirectionFlowchart}
 \end{figure*}

\begin{figure}[htbp]
 \centering
 \includegraphics[scale=0.5, trim={0cm 0cm 0cm 0cm},clip]{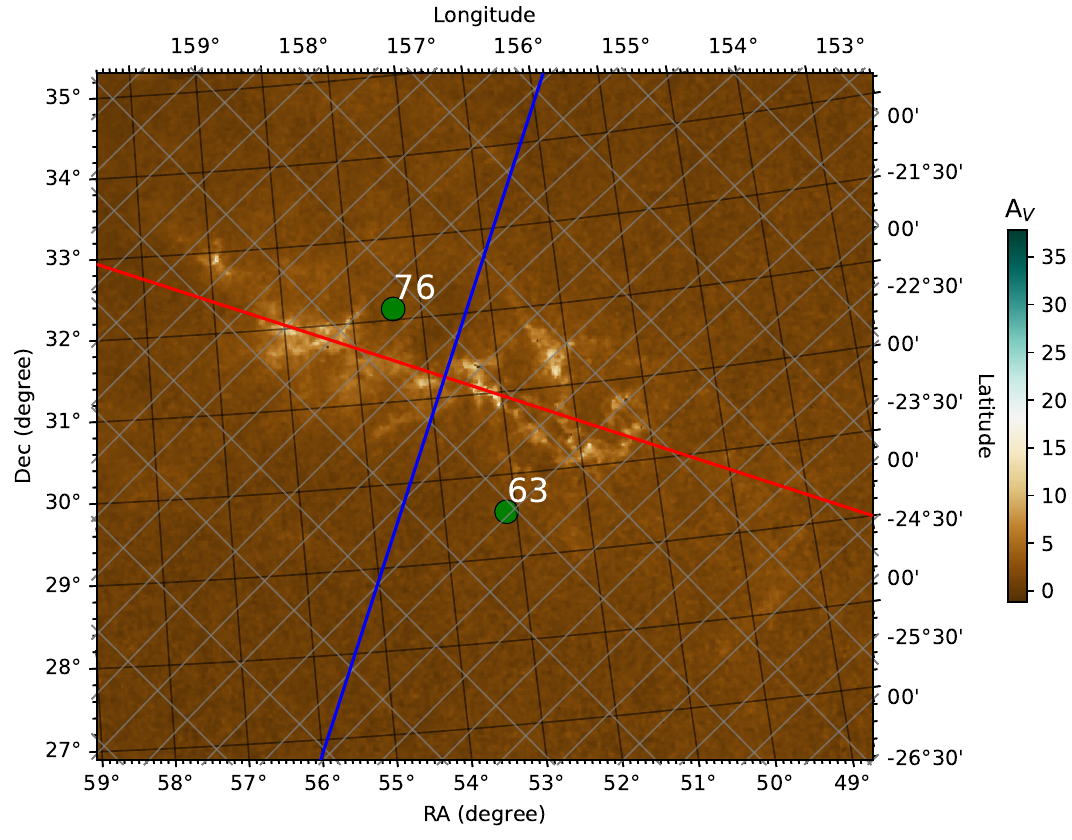}
 \caption{Perseus quadrants and anomalous (rejected) reference points (green circles) identified using the default input parameters for the Perseus Molecular Cloud. The background image shows the extinction map of the region. Points 63 and 76 are both too close to high extinction regions. No anomalous reference RM values were found. The cloud is divided into four regions for a thorough selection of reference points. If selected, the software ensures that each quadrant contributes equally to the determination of \rmref.} 
 \label{fig:PerseusAnomalous_Quadrant}
 \end{figure}

To find \rmref , the following steps are taken as illustrated in Figure (Flowchart)~\ref{flow:DirectionFlowchart}: 
\begin{enumerate}
    \item An OFF extinction threshold (\Avth) is set by default or by the user. \Avth\ determines the maximum \Av\ value that can be considered as a potential OFF position -- higher \Av\ values would indicate that their corresponding RM is influenced by the cloud's magnetic field. The software determines the appropriate \Avth\ based on the Galactic coordinates of each cloud, using the center location estimated from high extinction regions in the corresponding fits map. The default \Avth\ is as follows:
    \begin{enumerate}
        \item For high-latitude clouds (absolute central latitude $>15^{\circ}$), the default \Avth\ is 1. These clouds are relatively isolated from the complex Galactic plane, allowing for a lower threshold.
        \item For anti-galactic-pointing (pointing away from the Galactic center) and low-latitude clouds (absolute central latitude $<15^{\circ}$), the default \Avth\ is 1.5. These clouds experience moderate Galactic interference, necessitating a slightly higher threshold.
        \item For galactic-pointing (pointing towards the Galactic center) and low-latitude clouds, the default \Avth\ is 2. These clouds are in regions of high Galactic complexity, requiring the highest threshold to distinguish cloud effects from the Galactic contribution.
    \end{enumerate}
    \item All the RM points are sorted based on their \Av\ value, from lowest up to \Avth . This set of RM points defines the potential OFF points. 
    \item The software searches for anomalous OFF points in the potential OFF points set. The anomalous points include any potential OFF position that is too close to the cloud or has an RM value significantly different from the rest of the OFF positions.
    \begin{itemize} 
        \item Too close to the cloud: If the user has set ``\texttt{use near high extinction exclusion}" = True (default) in the config file, then any point too close to high extinction regions will be excluded from the potential OFF points. This is because points too close to the cloud might be influenced by the cloud's magnetic field. 
        The user can adjust the ``too close'' criterion by adjusting the ``\texttt{near high extinction threshold multiplier}" and ``\texttt{high extinction multiplier}" parameters in the config file. 
        The ``\texttt{high extinction threshold multiplier}" is multiplied by \Avth\ to set a high \Av\ value. Reference points should be located far enough away from regions with extinction values above this threshold.  
        The ``\texttt{near high extinction multiplier}" is multiplied by the resolution or pixel size of the extinction or column density map to define the proximity threshold. Points closer to high extinction regions than this threshold are considered too close and will be excluded from the potential OFF points. We discuss these choices in more details in Section~\ref{sec:Results}.
        \item Anomalous RM value:
        If the user has set ``\texttt{use anomalous value removal}" = True (default) in the config file, potential OFF points with RM values that are significantly higher or lower compared to other OFF positions are excluded from the final reference points. These outlier RM values are likely caused by astrophysical phenomena other than the Galactic contribution to the RM, such as intrinsic RM from the source itself. 
        The user can adjust the definition of an ``anomalous RM'' by modifying the ``\texttt{anomalous values iqr multiple (greater than or equal to)}" ($C_{IQR , RM}$) in the config file. The software calculates the median ($RM_{med}$) and interquartile range ($RM_{IQR}$) of all potential OFF points. The interquartile range is the difference between the third quartile (Q3) and the first quartile (Q1). If a potential OFF point has an RM value higher than $RM_{\text{Upper Limit}}= RM_{med} + C_{IQR , RM} \times RM_{IQR}$ or lower than $RM_{\text{Lower Limit}}= RM_{med} - C_{IQR , RM} \times RM_{IQR}$, it is considered anomalous.  
        \item Too far from the cloud: This condition is similar to the ``too-close-to-cloud" condition. If the user has set ``\texttt{use far high extinction exclusion}" = True in the config file, then any point too far from the cloud or high extinction regions will be excluded from the potential OFF points set. This is necessary to ensure that we are sampling the Galactic contribution to the RM (for subtraction from \rmon ), requiring a similar path and path length along the line of sight. The software's default setting is False, as the user already identifies the desired region for the RMs (ON and OFF) to be considered in the config file or through the cloud column density or extinction fits file. 
        The user can adjust the definition of ``too far" by modifying the ``\texttt{far from high extinction threshold multiplier}".  The ``\texttt{far from high extinction multiplier}" parameter is multiplied by the resolution or the pixel size of the extinction (or column density) map of the cloud to define the distance threshold. 
        The ``\texttt{high extinction threshold multiplier}" is  used to set an \Av\ value, and RM points should not be too far from this value. Figure~\ref{fig:PerseusAnomalous_Quadrant} illustrates anomalous potential OFF points identified using the default input parameters for the Perseus molecular cloud. 
    \end{itemize} 
    \item The stability trend is used to determine the optimal number of candidate OFF points at which the trend of calculated \blos\ values stabilizes. This is done by analyzing the differences in \blos\ values for different numbers of OFF points and identifying the lowest number of OFF points needed for a stable \blos\ behavior  
    when more OFF points are added. The goal is to find the lowest number of reference points that result in a stable trend of \blos\ values, avoiding the use of too many OFF points in a limited RM dataset and allowing determination of \blos\ associated with lower density regions.  \citet{Tahanietal2018} determined the optimal number of reference points  primarily through visual inspection of stability trend plots (see their Figure 4), and this process is now automated. Figure~\ref{fig:Stability} illustrates the stability trend plot for the Perseus cloud. The software's approach to the stability trend is as follows:
        \begin{itemize}
            \item[$\blacksquare$] Initially, all non-anomalous potential OFF points are gathered. The software calculates the \blos\ values using an increasing number of OFF points ordered by extinction value from low to high.  That is, the software starts with a single OFF point with the lowest extinction value, then the two OFF points with the lowest extinction values, and so on, up to including all N potential OFF points. This process generates a set of \blos\ values for each RM point, enabling the exploration of the effects of reference points and the determination of the optimal number of OFF points. 
            \item[$\blacksquare$] For each set of OFF points, the differences between adjacent \blos\ values (calculated using i and i+1 reference points) are computed for the remaining RM points. 
            The maximum and minimum \blos\ differences among all the RM points are considered the stability maximum and minimum thresholds. These thresholds enable the determination of when the \blos\ values stabilize as OFF points are added. The software determines the number of OFF points at which most RM points have adjacent \blos\ differences within the stability thresholds, and this number is taken as the optimal number of reference points. 
        \end{itemize}
    \item If the user has set the quadrant selection to True, the software divides the cloud into four quadrants based on extinction values of the region\footnote{To divide the cloud into quadrants, first a center point in the map is found weighted by extinction distribution of the region. Then the software finds a  line passing through this center point and divides the entire region into two equally-weighted regions. A perpendicular line passing through the first line and the center point is then found, dividing the cloud into four quadrants.}. The software starts taking points one by one as sorted in the potential OFF points set until it satisfies the minimum number of points per quadrant specified by the user. If there are not enough potential OFF points in a quadrant, the software continues taking  OFF points from the sorted set until it takes all the  OFF points in that quadrant. Figure~\ref{fig:PerseusAnomalous_Quadrant} illustrates the quadrants of the Perseus cloud. 
    \item After the final OFF (reference) points are selected, an average is calculated to determine \rmref. The software default is to simply average the RM value of all the selected reference points to determine \rmref . 
    \begin{equation}
    \begin{aligned}
    \text{RM}_{\text{ref}} = \sum_{i=1}^N \frac{\text{RM}_{\text{OFF, i}}}{N}\\
    \text{A}_{\text{V, ref}} = \sum_{i=1}^N \frac{\text{A}_\text{{V~OFF, i}}}{\text{N}}.
    \end{aligned}
    \end{equation}
    However, if the user has set the quadrant weighting scheme to True, a weighting scheme is applied to ensure that the effects of each cloud quadrant are the same in finding \rmref. This ensures that if, for example, different quadrants have different numbers of OFF points, they both contribute equally to \rmref .
    \item After \rmref\ is found, the contribution of the molecular cloud's magnetic field to each \rmon\ can be estimated as follows:
    \begin{equation}
    \begin{aligned}
    \text{RM}_{\text{MC}} = \text{RM}_{\text{ON}} - \text{RM}_{\text{ref}} =   \bigg( 0.812 \int n_{\text{e}} B_{\text{LOS}}dl \bigg)_{\text{MC}},
    \end{aligned}
    \end{equation}
    and the extinction value of the cloud at each ON point is:
    \begin{equation}
        A_{\text{V, MC}} = A_{\text{V, ON}} - A_{\text{V, ref}}.
    \end{equation}
\end{enumerate}
In summary, this process for determining \rmref\ is designed to be both robust and flexible. It accounts for various factors that could influence the accuracy of the reference RM, including proximity to high-extinction regions, anomalous RM values, and the spatial distribution of reference points across the cloud. The stability trend analysis ensures that an optimal number of reference points is used, balancing the need for statistical reliability with the preservation of data for \blos\ determination.   This comprehensive approach to \rmref\ determination forms a crucial foundation for accurate \blos\ mapping in molecular clouds using the MC-BLOS technique.


\begin{figure}[htbp]
 \centering
 \includegraphics[scale=0.6, trim={0cm 0cm 0cm 0cm},clip]{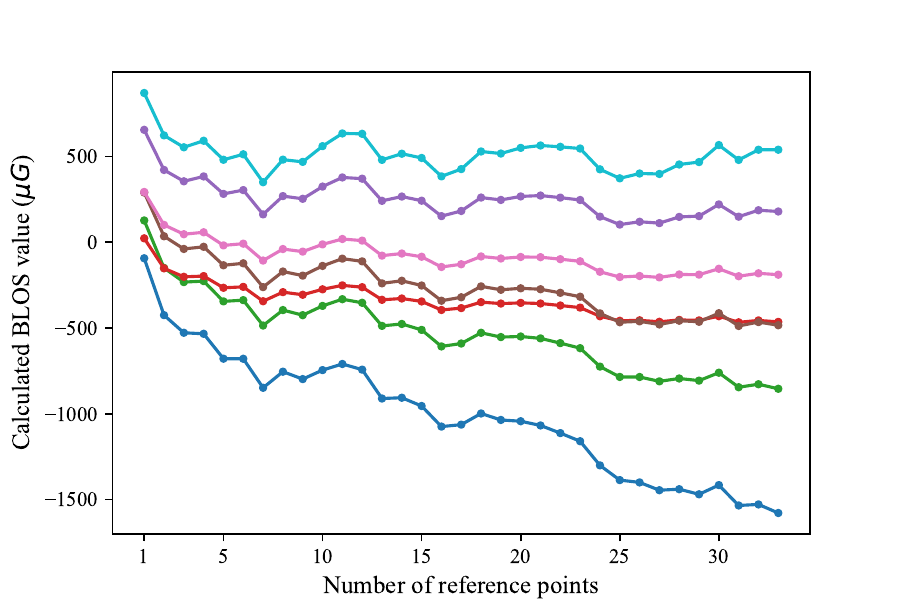}
 \caption{Stability trend plot for the Perseus cloud, showing the number of reference points (x-axis) versus the calculated \blos\ values (y-axis) for each RM point. The \blos\ of each RM point is calculated by taking an increasing number of OFF points to find the optimal number.} 
 \label{fig:Stability}
 \end{figure} 

\subsection{Strength determination}
\label{sec:strengthDetermin}
After \rmref\ and \Avref\ are determined, the \blos\ of each remaining ON point can be obtained using \Av\ (or column density) maps and a chemical evolution code. The software reads and analyzes chemical results from output files generated by a previously tested chemical evolution code~\citep{Gibsonetal2009}. These results cover various cloud parameters: density ($n$), temperature ($T$), cosmic ionization rate ($IR_{CR}$), and ultra-violet ionization rate ($IR_{UV}$). Users can also employ their own chemical evolution code (see the user manual for details). The current chemical code divides the cloud into layers and calculates the extinction and element abundances (including electrons) for each layer.

Concurrently, the \Av\ (or column density) of each point is read from the maps. Assuming a symmetrical cloud, the electron column density is calculated for each layer up to the cloud's center along the line of sight, reaching $\frac{A_{V, MC}}{2}$, as illustrated in Figure~\ref{fig:cloudLayerCartoon}. The total electron column density for each ON point is then obtained by doubling this value, as explained in Section~\ref{sec:findingNe}. Finally, the \blos\ strength is determined using equations \ref{eq:findBlos}, \ref{eq:FinalNe}, and \ref{eq:RMMC}.
\begin{figure*}
    \centering
    \scalebox{0.85}{
        \begin{tikzpicture}[node distance=2cm]

    \node (init) [invisible] {};
    \node (start) [begin, below of =init, xshift = 10cm] {$N_e$ Determination};
    
    \node (select_results) [process, below of=start, text width = 4cm, yshift = -1.3cm, xshift = -5cm] {Retrieve chemical results using corresponding input parameters ($IR_{CR}$, $IR_{UV}$, $T$, $n$) };

     \node (find_results) [process, below of=start, text width = 4cm, yshift = -1.1cm, xshift = 5cm] {Find corresponding $A_V$ (or $N$) at each $RM_{ON}$ };
    
    \node (find_ar_n) [process, below of=select_results, text width=4cm, yshift = -0.6 cm] {Divide cloud into layers at each ON point to $\frac{A_V}{2}$ and find $X_{e,i}$ in each layer};
  
    \node (find_nei) [process, below of=find_ar_n, text width=4cm, yshift = -0.3cm] {Determine N$e,i$ \\ based on $X_{e,i}$  and $A_{v,i}$};
    
    \node (add_nei) [startstop, below of=find_nei, text width=4cm, yshift = 0.1cm] {Find $2 \times \Sigma_i N_{e,i}$ };

    \draw [arrow](start) -- ++(0,-2) -- ++(-5,0)--++(0,-0.6);
    \draw [arrow](start) -- ++(0,-2) -- ++(5,0)--++(0,-0.5);
    \draw [arrow] (select_results) -- (find_ar_n);
    \draw [arrow] (find_ar_n) -- (find_nei);
    \draw [arrow] (find_nei) -- (add_nei);
    \draw [arrow](find_results) -- ++(0,-1.7) -- ++(-9.95,0);
    
\end{tikzpicture}}
    \caption{Flowchart illustrating the strength determination steps. In this process, a chemical evolution code along with extinction maps are used to estimate the electron column densities at each layer of the cloud. The output results of a chemical code are included in the software for different cloud density ($n$), temperature ($T$), cosmic ionization rate ($IR_{CR}$), and ultra-violet ionization rate ($IR_{UV}$). }
 \end{figure*}
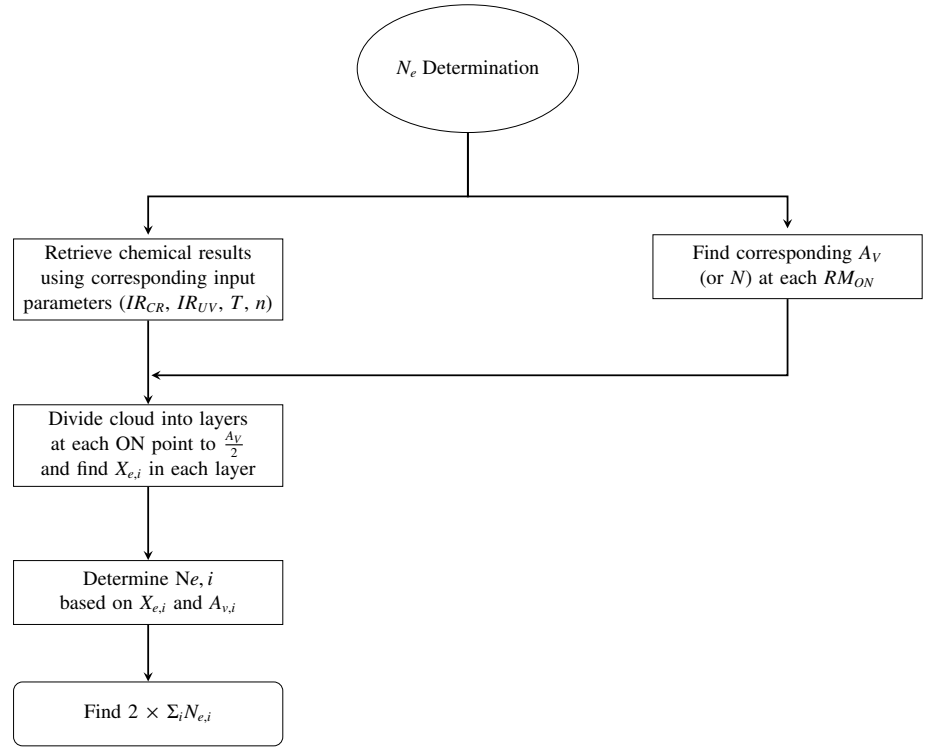

\subsection{Finding total uncertainty}
\label{sec:uncertaintyUpgrade}

As discussed in Section~\ref{sec:uncertainty}, the uncertainties of each point are obtained by considering the uncertainties in RM, \rmref , electron column densities, and the overlap of RM and \Av . The electron column density uncertainties stem from uncertainties in extinction and the chemical evolution code. We calculate the \Ne\ uncertainty by varying the chemical code's input temperature by $\pm 20\%$ and input volume density by $\pm 50\%$, as well as finding the maximum and minimum \Av\ values within a Jeans length radius around the RM value.

The uncertainties in RM and \rmref\ can cause a change in the direction of the magnetic field, as these parameters determine the direction of each \blos\ point. Conversely, the \Ne\ uncertainties only affect the strength of \blos, not its direction. The direction is determined solely by RM and \rmref, as the sign of \rmon\ determines the direction. 
The upgraded technique now accounts for this distinction; if the \Ne\ uncertainties exceed the \blos\ value, the software adjusts these uncertainties to ensure they do not affect the sign of \blos .

Additionally, the software ensures all ON points have an \Av\ value that is a user-specified multiple higher than \Avref\ (i.e., $A_{{\rm{V, ON}}} \geq \text{multiple} \times A_{\rm{V,ref}}$). This criterion enhances the fidelity of magnetic field strength determination, as the \Av\ value is crucial for strength determination. The multiplication factor for setting the minimum ON point extinction can be adjusted in the input parameter file by changing the ``\texttt{on point extinction multiple of off point average multiplier}". For example, if set to three, the software ensures all ON points have $A_V \geq 3 \times A_{\rm{V, ref}}$. While ideally this factor should be greater than one, the current default is set to one due to the low source density in available RM catalogs. Future RM catalogs with higher source densities will enable better selection of ON points.

Regardless of the chosen multiplication factor, the software always ensures that any ON point has an \Av\ value higher than each of the reference points. This criterion does not influence the \blos\ direction. Even with a multiplication factor of one, we obtain \blos\ directions that match previously determined directions from Zeeman observations~\citep{Heiles1987, Heiles1997}. 

While the \blos\ strengths of points with \Av\ not significantly higher than \Avref\ may exceed those from Zeeman observations, they show good correspondence with Zeeman-obtained strengths. For example, \citet{Tahanietal2018} found that the error-weighted average \blos\ on the western side of Orion A is half that of its eastern side, consistent with Zeeman observations \citep{Tahanietal2022O}.

\section{Results and Discussion}
\label{sec:Results}
In this section, 
we present the MC-BLOS final results for the Orion A, Perseus, California, and Orion B molecular clouds and compare them with the previously-published results by \citet{Tahanietal2018} and nearby molecular Zeeman effect observations. Although users can influence the selection of reference points, the software-suggested reference points produce results consistent with \citet{Tahanietal2018}, with differences in the final \blos\ values falling within the uncertainty ranges for each cloud.

To compare our results with those of \citet{Tahanietal2018}, we use the RM catalog of \citet{Tayloretal2009}. As the same RM catalogs are used in both studies, the main differences in final \blos\ values should be due to differences in \rmref\ values. We examine the software-selected \rmref\ using three input parameter sets:
\begin{enumerate}
    \item Default input parameters for each cloud (``\texttt{Default}" parameter set). 
    \item Default input with at least one reference point in each cloud quadrant (``\texttt{Min1PerQ}" parameter set). 
    \item Default input with the ``\texttt{high extinction threshold multiplier}" increased from 2 to 5, positioning OFF points further from high extinction regions (``\texttt{NearExt5}" parameter set). 
\end{enumerate}

Table~\ref{table:rmref} compares the \rmref\ values generated using these parameter sets with those of \citet{Tahanietal2018}. The software-selected reference points in all three parameter sets result in \rmref\ values similar to those of \citet{Tahanietal2018}, falling within their uncertainty ranges. For Orion A and California, while the \rmref\ values deviate from the original \citet{Tahanietal2018}, they remain consistent within the uncertainties.

Although the \rmref\ values of Orion A slightly deviate from the original ones, the direction of the reversal seen in the Orion A cloud maintains the same orientation as \citet{Tahanietal2018} and Zeeman-obtained directions~\citep{Heiles1997}. In both Orion A and Orion B clouds, we find points that closely align with previously published molecular Zeeman observations.

In the Orion A region near $\alpha(J2000) \simeq 83.81^{\circ}, \delta(J2000) \simeq -5.37^{\circ}$, different molecular Zeeman values have been reported, such as $+360 \pm 80,\mu$G \citep{Falgaroneetal2008, Crutcher1999APJ520, Crutcheretal2010}, $-79 \pm 99,\mu$G \citep{Crutcheretal1996}, and $-40 \pm 240,\mu$G \citep{Crutcher1999APJ514, Crutcheretal2010}. 
Considering the uncertainties, these may indicate magnetic field strengths ranging from $+440,\mu$G to $-280,\mu$G. 
Near this point, similar to the Zeeman measurements, we find high uncertainties and magnetic field values of $-45 \pm 36,\mu$G and $-6 \pm 35,\mu$G at $\alpha(J2000) = 83.90^{\circ}, \delta(J2000) = -5.39^{\circ}$ and $\alpha(J2000) = 83.90^{\circ}, \delta(J2000) = -5.38^{\circ}$, respectively.

In Orion B, the observations are near $\alpha(J2000) \simeq 85.44^{\circ}, \delta(J2000) \simeq 1.93^{\circ}$ with the values of $-270 \pm 330,\mu$G \citep{Crutcher1999APJ514} and $-87 \pm 5.5,\mu$G \citep{Crutcheretal1999APJ515}. Near this location, at $\alpha(J2000) = 85.42^{\circ}, \delta(J2000) = 1.92^{\circ}$ and $\alpha(J2000) = 85.42^{\circ}, \delta(J2000) = 1.90^{\circ}$, we find values of $-110_{-24}^{+24}\,\mu$G and $-119_{-27}^{+28}\,\mu$G, respectively, with the same direction.

We note that in high extinction regions, our \blos\ values may be smaller than expected. This is because our method considers all cloud layers along the line of sight, including potential depolarization effects. Depolarization can occur due to variations in magnetic field direction across different cloud layers. Changes in the sign of the magnetic field along the line of sight can lead to a reduction in the net observed \blos\ value. This integrated approach may result in lower apparent field strengths in regions where the magnetic field structure is more complex or tangled.

\noindent
\begin{table*}
\centering
\label{table:rmref}
\renewcommand{\arraystretch}{1.5}  
\begin{tabular}{|c||*{5}{c|}}\hline
\backslashbox{Cloud}{Model}
&\makebox[4.3em]{\texttt{Default}}&\makebox[4.3em]{\texttt{Min1PerQ}}& \makebox[4.3em]{\texttt{NearExt5}}& \makebox[4.3em]{Tahani+18}\\\hline\hline
Orion A &$15.2 \pm 12.6$& $15.2 \pm 12.6$& $8.7 \pm 13.0$& $1.4 \pm 13.7$\\\hline
Perseus &$34.1 \pm 10.6$&$34.1 \pm 10.6$& $38.7 \pm 12.7$& $31.1 \pm 11$\\\hline
California &$13.4 \pm 12.9$&$13.4 \pm 12.9$& $13.4 \pm 12.9$& $4.0 \pm 14$\\\hline
Orion B &$26.6 \pm 9.6$&$31.3 \pm 11.7$& $26.6 \pm 9.6$& $32.3 \pm 10$\\\hline
\end{tabular}
\hspace{10pt}
\caption{\rmref\ values for each cloud and model using the upgraded technique and \rmref\ from \citet{Tahanietal2018}. The differences between the models for each cloud are within RM uncertainties.}
\end{table*}

\begin{figure*}[htbp]
 \centering
 \includegraphics[scale=0.34, trim={0cm 0cm 0cm 2cm},clip]{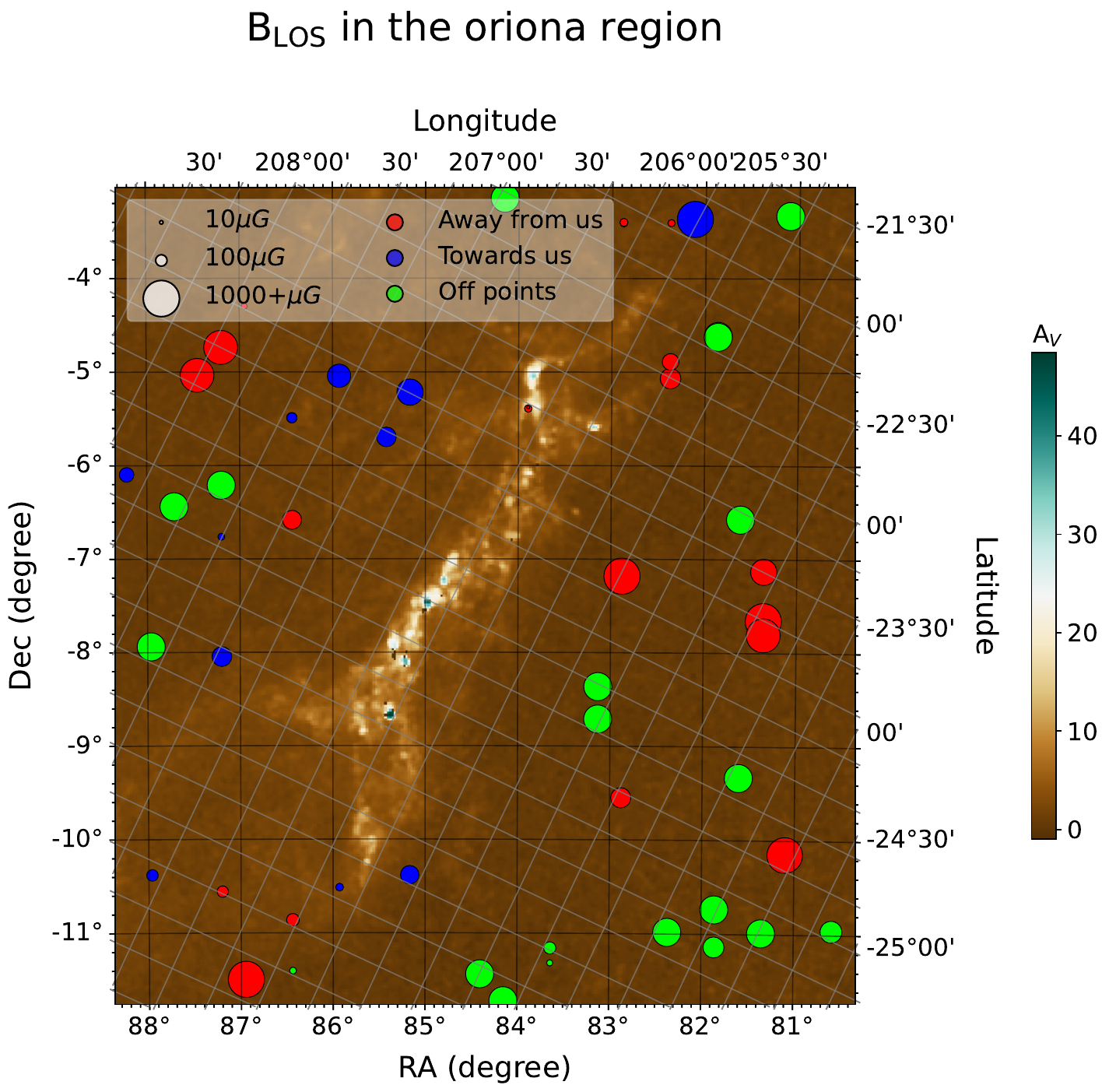}
 \includegraphics[scale=0.34, trim={0cm 0cm 0cm 2cm},clip]{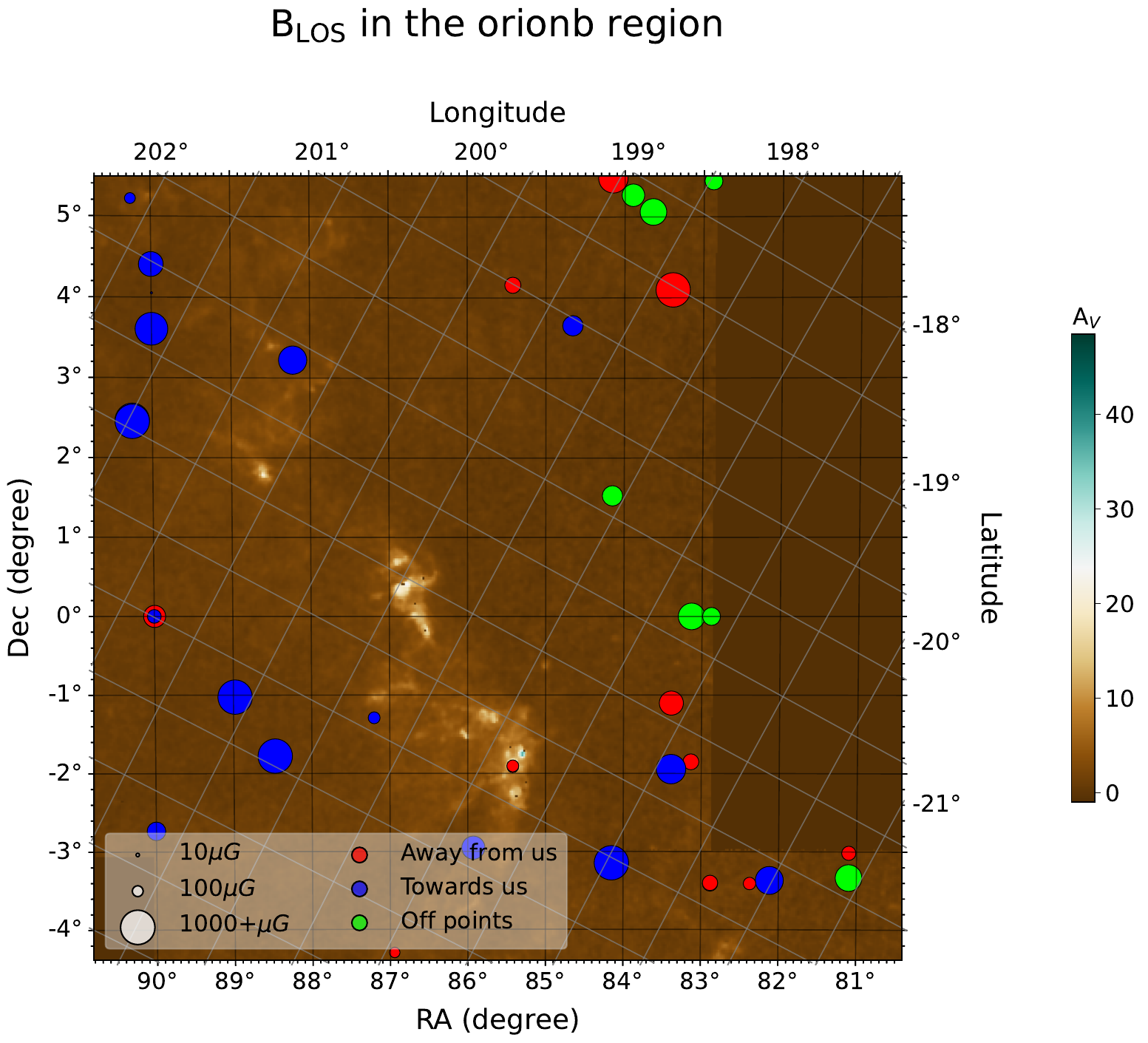}\\
 \includegraphics[scale=0.325, trim={0cm 0cm 0cm 2cm},clip]{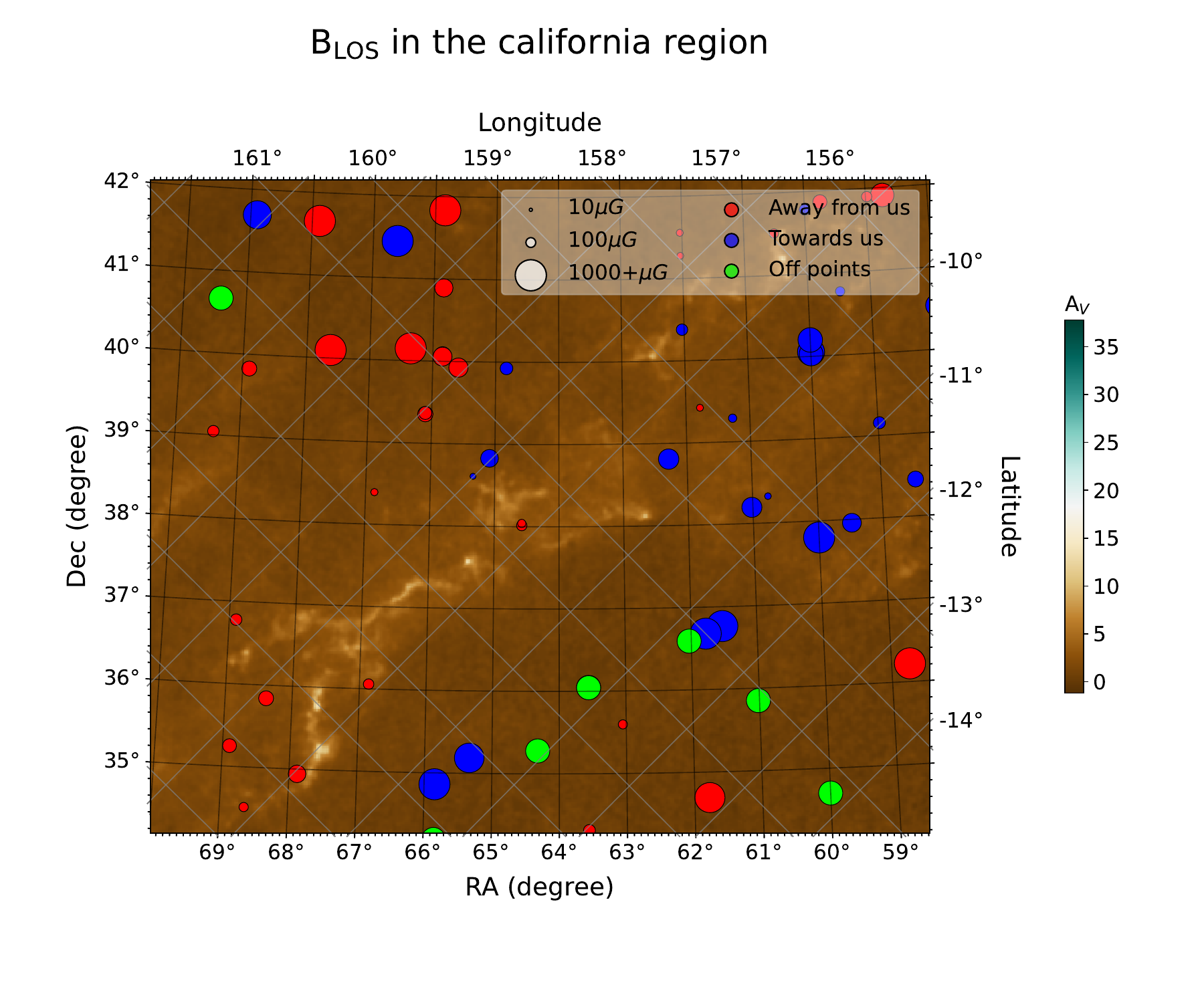}
 \includegraphics[scale=0.325, trim={0cm 0cm 0cm 2cm},clip]{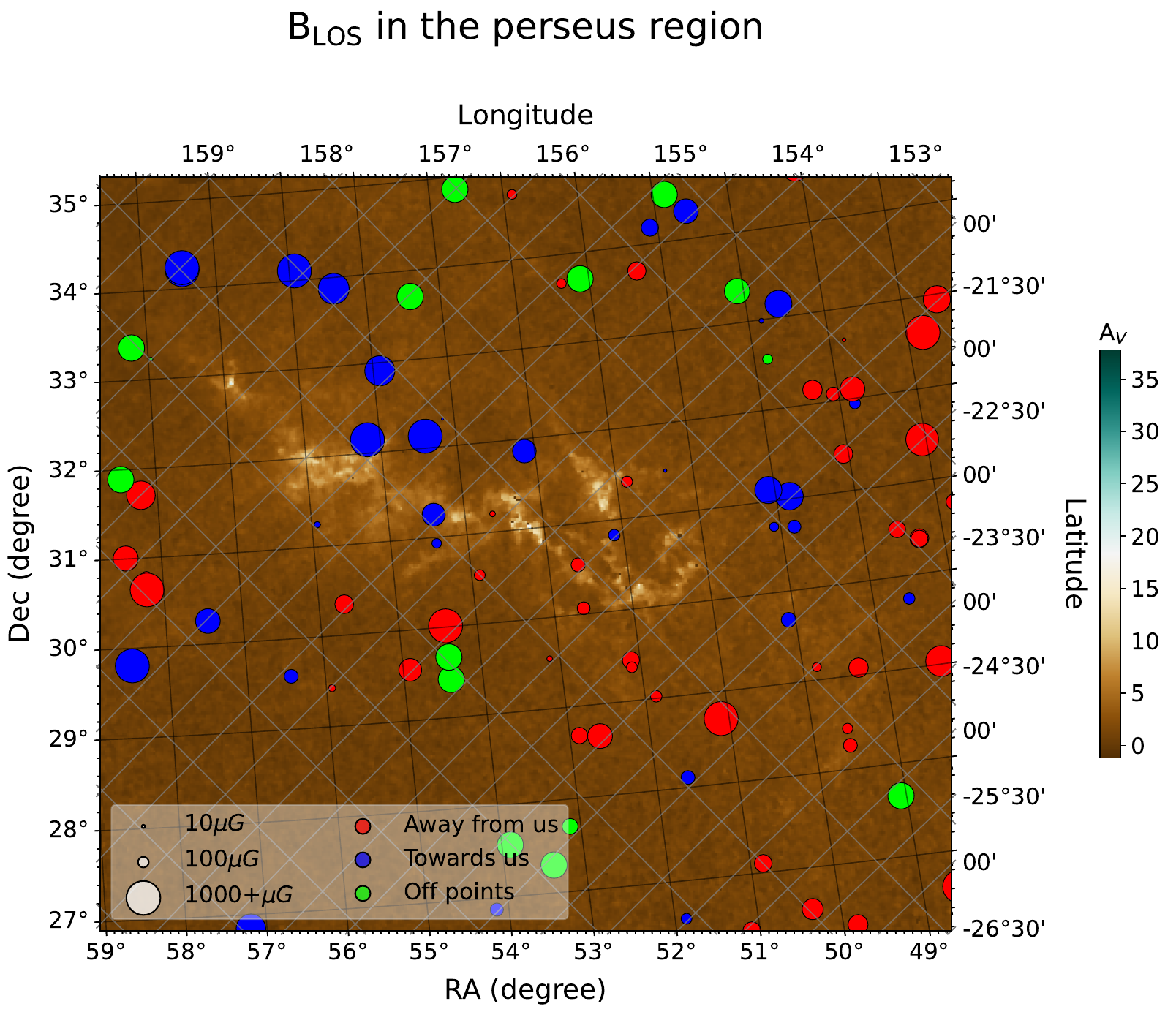}
 \caption{Final \blos\ maps of the Orion A, Orion B, California, and Perseus molecular clouds using \texttt{Default} parameter set. The background color image shows extinction, and the circles represent \blos\ strengths (without uncertainties). Red (blue) circles indicate magnetic fields pointing away from (toward) the observer.} 
 \label{fig:FinalResultsAllClouds}
 \end{figure*} 

\begin{figure}[htbp]
 \centering
 \includegraphics[scale=0.34, trim={0cm 0cm 0cm 2cm},clip]{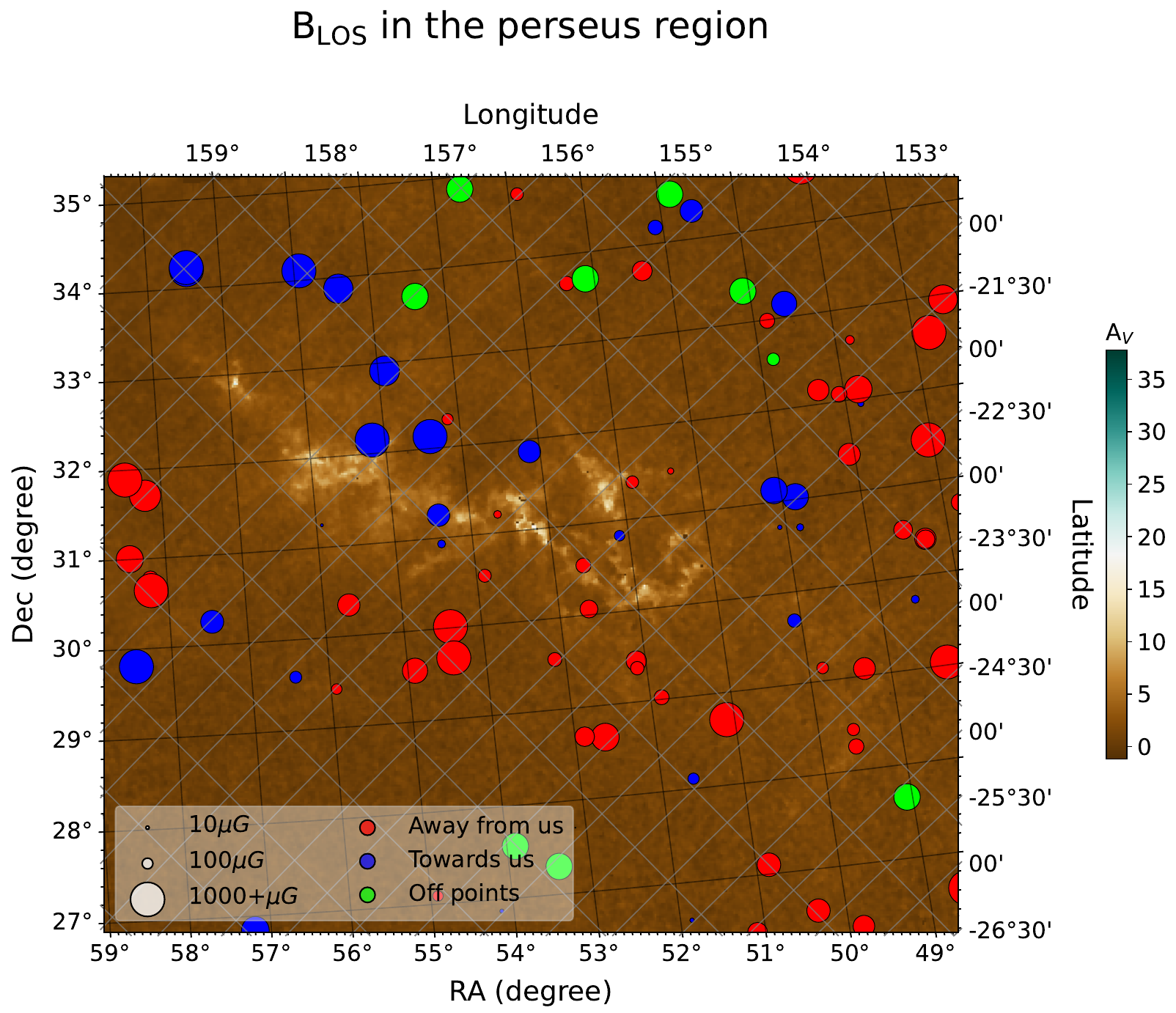}
 \caption{Final \blos\ maps of the Perseus molecular clouds using \texttt{NearExt5} parameter set, positioning the reference point (green circles) further from the cloud. The background color image shows extinction, and the circles represent \blos\ strengths (without uncertainties). Red (blue) circles indicate magnetic fields pointing away from (toward) the observer.} 
 \label{fig:PerseusNearExt5}
 \end{figure} 

Figure~\ref{fig:FinalResultsAllClouds} illustrates the final \blos\ maps of the four clouds. The background image shows extinction, and the circles depict \blos\ direction and strength (without uncertainty values). Red, blue, and green circles illustrate magnetic fields pointing away from us, toward us, and the reference points, respectively.

As mentioned earlier, the software's default multiplication factor for setting the minimum accepted \Avon\ value is currently set to one. 
This can lead to ON and reference points being positioned near each other. Despite this, we still obtain acceptable \blos\ results because having a low \Av\ value is the most important criterion for setting OFF positions due to the clumpiness of the clouds. The abundance of RM sources in upcoming catalogs will result in better separation of ON and OFF points. As shown in Figure~\ref{fig:PerseusNearExt5}, this can also be addressed by using the \texttt{NearExt5} parameter set, which positions OFF points further from the cloud. We note that both the \texttt{NearExt5} and \texttt{Default} parameter sets produce very similar results.

Figure~\ref{fig:FinalResultsAllClouds} exhibits the same \blos\ morphology as \citet{Tahanietal2018}. The Orion A, Perseus, and California molecular clouds show a reversal of \blos\ from one side of the cloud to the other (along the cloud's minor axis). 
The RM points associated with Orion B are insufficient for a field reversal conclusion. This reversal in Orion A with the same orientation was previously observed using atomic Zeeman observations~\citep{Heiles1997, PressRelease}.

\citet{Tahanietal2022P} reconstructed the complete 3D magnetic field of the Perseus cloud (ignoring sub-parsec fluctuations along the field lines). 
The reconstructed 3D magnetic field of the Perseus cloud by \citet{Tahanietal2022P} suggested the presence of an interacting structure near the cloud, which was responsible for the observed \blos\ reversal. The presence of proposed structure was later supported by kinematic observations presented in \citet{Kounkeletal2022}, providing independent confirmation of the proposed interaction. 
The agreement between the implications of the reconstructed 3D magnetic field structure and subsequent kinematic observations strengthens the reliability and validity of both the 3D fields (obtained using the MC-BLOS results) and the MC-BLOS method itself.

We note that the uncertainties in \rmref\ in our study and \citet{Tahanietal2018} are expected to be significantly reduced by upcoming observations, such as those with the Square Kilometer Array Observatory, which will provide more accurate and abundant RM data. This should improve the overall accuracy and precision of the MC-BLOS results.

\section{Summary and Plans for maintaining MC-BLOS}
\label{Sec:summary}
\subsection{Summary}

We have upgraded the Faraday-based technique of \citet{Tahanietal2018} and present MC-BLOS (v1.0), a Python software package for determining the line-of-sight component of magnetic fields associated with molecular clouds. 
The software automates the process of determining reference points, which previously required manual analysis, and incorporates upgrades in the selection and handling of OFF points and \blos\ uncertainty values. MC-BLOS utilizes available rotation measure catalogs (point sources), extinction (or column density) maps, and output results from chemical evolution modeling to produce \blos\ maps and tables associated with molecular clouds.

The software employs an ON-OFF approach to find the RM contribution of the cloud. By using a number of OFF points that do not probe the magnetic field of the clouds, the contribution of everything along the line of sight except the cloud and its magnetic field can be determined. 
The sign of RMs at each point associated with the cloud determines the \blos\ direction. To determine the strength of magnetic fields, the software uses the RM induced by the cloud at each point (determined using \rmref\ and the RM value at each point) and electron column densities, which are obtained for each point using the extinction (column density) maps and the chemical evolution code.

The results presented in this paper incorporated the previously available RM catalog of \citet{Tayloretal2009}. This catalog used the VLA Sky Survey data \citep[NVSS;][]{Condon1998}, which had only two observing frequencies, resulting in high uncertainties when inferring RM values.   
Upcoming RM observations by POSSUM, VLASS, and SKA are expected to significantly improve the RM uncertainties and increase the RM density by a factor of 10 \citep[e.g.,][]{Vanderwoudeetal2024}, resulting in improved \rmref\ uncertainties. Using the MC-BLOS software along with these future observations, the line-of-sight magnetic field component of many molecular clouds can be determined with greater accuracy and precision.

\subsection{Plans for maintaining and upgrading MC-BLOS software}

As this paper is the first official publication of MC-BLOS software, we are committed to its continuous improvement and upgrade. Our development roadmap is guided by new and improved observations and feedback from the scientific community. For instance, recent VLA observations (observation IDs VLA/19B-053 and VLA/24A-376) not only enable detailed \blos\ and 3D magnetic field maps of the Perseus cloud (Hajizadeh et al., in prep.) but also provide an opportunity to refine the \rmref\ determination in MC-BLOS, thanks to their increased source density compared to the NVSS catalog.

we envision several key areas for improvement: 
\begin{itemize}
    \item Improved \rmref\ determination: With higher source densities and more OFF points, we plan to replace the current stability trend method with more advanced techniques. This includes:
    \begin{itemize}
        \item Implementing sophisticated statistical methods for robust outlier detection and handling.
        \item Exploring the potential of machine learning algorithms to recognize patterns in RM distributions and optimize reference point selection.
        \item Utilizing 3D dust maps.
    \end{itemize}
    These enhancements aim to improve the accuracy and reliability of \rmref\ determinations, which are crucial for precise \blos\ mapping.
    \item Algorithmic Improvements: We plan to refine our algorithms as more RM data become available, particularly in areas such as anomaly detection, cloud boundary determination, and the handling of complex cloud morphologies. 
\end{itemize}

To facilitate these advancements, we encourage feedback, bug reports, suggestions, and contributions from the scientific community. We plan to maintain and update the software regularly, releasing new versions as significant improvements are made.

\section*{acknowledgments}
M.Tahani is supported by the Banting Fellowship (Natural Sciences and Engineering Research Council Canada) hosted at Stanford University and the Kavli Institute for Particle Astrophysics and Cosmology (KIPAC) Fellowship.  Claude.AI and QuillBot\footnote{\url{https://quillbot.com/}} were used for proofreading and language refinement. We have used PyCharm. 

\software{Astropy~\citep{astropy},
Matplotlib \citep{Matplotlib}, Math \citep{Mathpython}, Numpy \citep{Numpy}, Pandas \citep{Pandas},  Scikit-learn \citep{sklearn}.}

\bibliography{Allbiblio}{}

\begin{thebibliography}{}
\expandafter\ifx\csname natexlab\endcsname\relax\def\natexlab#1{#1}\fi
\providecommand{\url}[1]{\href{#1}{#1}}
\providecommand{\dodoi}[1]{doi:~\href{http://doi.org/#1}{\nolinkurl{#1}}}
\providecommand{\doeprint}[1]{\href{http://ascl.net/#1}{\nolinkurl{http://ascl.net/#1}}}
\providecommand{\doarXiv}[1]{\href{https://arxiv.org/abs/#1}{\nolinkurl{https://arxiv.org/abs/#1}}}

\bibitem[{{Arzoumanian} {et~al.}(2021){Arzoumanian}, {Furuya}, {Hasegawa}, {Tahani}, {Sadavoy}, {Hull}, {Johnstone}, {Koch}, {Inutsuka}, {Doi}, {Hoang}, {Onaka}, {Iwasaki}, {Shimajiri}, {Inoue}, {Peretto}, {Andr{\'e}}, {Bastien}, {Berry}, {Chen}, {Di Francesco}, {Eswaraiah}, {Fanciullo}, {Fissel}, {Hwang}, {Kang}, {Kim}, {Kim}, {Kirchschlager}, {Kwon}, {Lee}, {Liu}, {Lyo}, {Pattle}, {Soam}, {Tang}, {Whitworth}, {Ching}, {Coud{\'e}}, {Wang}, {Ward-Thompson}, {Lai}, {Qiu}, {Bourke}, {Byun}, {Chen}, {Chen}, {Chen}, {Cho}, {Choi}, {Choi}, {Chrysostomou}, {Chung}, {Dai}, {Diep}, {Duan}, {Duan}, {Eden}, {Fiege}, {Franzmann}, {Friberg}, {Fuller}, {Gledhill}, {Graves}, {Greaves}, {Griffin}, {Gu}, {Han}, {Hatchell}, {Hayashi}, {Houde}, {Jeong}, {Kang}, {Kang}, {Kataoka}, {Kawabata}, {Kemper}, {Kim}, {Kim}, {Kim}, {Kim}, {Kirk}, {Kobayashi}, {K{\"o}nyves}, {Kusune}, {Kwon}, {Lacaille}, {Law}, {Lee}, {Lee}, {Lee}, {Lee}, {Lee}, {Li}, {Li}, {Li}, {Liu}, {Liu}, {Liu}, {Lu}, {Mairs}, {Matsumura}, {Matthews},
  {Moriarty-Schieven}, {Nagata}, {Nakamura}, {Nakanishi}, {Ngoc}, {Ohashi}, {Park}, {Parsons}, {Pyo}, {Qian}, {Rao}, {Rawlings}, {Rawlings}, {Retter}, {Richer}, {Rigby}, {Saito}, {Savini}, {Scaife}, {Seta}, {Shinnaga}, {Tamura}, {Tang}, {Tomisaka}, {Tram}, {Tsukamoto}, {Viti}, {Wang}, {Xie}, {Yen}, {Yoo}, {Yuan}, {Yun}, {Zenko}, {Zhang}, {Zhang}, {Zhang}, {Zhou}, {Zhu}, {de Looze}, {Dowell}, {Eyres}, {Falle}, {Friesen}, {Robitaille}, \& {van Loo}}]{Arzoumanianetal2021}
{Arzoumanian}, D., {Furuya}, R.~S., {Hasegawa}, T., {et~al.} 2021, \aap, 647, A78, \dodoi{10.1051/0004-6361/202038624}

\bibitem[{{Astropy Collaboration} {et~al.}(2013){Astropy Collaboration}, {Robitaille}, {Tollerud}, {Greenfield}, {Droettboom}, {Bray}, {Aldcroft}, {Davis}, {Ginsburg}, {Price-Whelan}, {Kerzendorf}, {Conley}, {Crighton}, {Barbary}, {Muna}, {Ferguson}, {Grollier}, {Parikh}, {Nair}, {Unther}, {Deil}, {Woillez}, {Conseil}, {Kramer}, {Turner}, {Singer}, {Fox}, {Weaver}, {Zabalza}, {Edwards}, {Azalee Bostroem}, {Burke}, {Casey}, {Crawford}, {Dencheva}, {Ely}, {Jenness}, {Labrie}, {Lim}, {Pierfederici}, {Pontzen}, {Ptak}, {Refsdal}, {Servillat}, \& {Streicher}}]{astropy}
{Astropy Collaboration}, {Robitaille}, T.~P., {Tollerud}, E.~J., {et~al.} 2013, \aap, 558, A33, \dodoi{10.1051/0004-6361/201322068}

\bibitem[{{Brown} \& {Taylor}(2001)}]{BrownTaylor2001}
{Brown}, J.~C., \& {Taylor}, A.~R. 2001, \apjl, 563, L31, \dodoi{10.1086/338358}

\bibitem[{{Ching} {et~al.}(2022){Ching}, {Qiu}, {Li}, {Ren}, {Lai}, {Berry}, {Pattle}, {Furuya}, {Ward-Thompson}, {Johnstone}, {Koch}, {Lee}, {Hoang}, {Hasegawa}, {Kwon}, {Bastien}, {Eswaraiah}, {Wang}, {Kim}, {Hwang}, {Soam}, {Lyo}, {Liu}, {Le Gouellec}, {Arzoumanian}, {Whitworth}, {Di Francesco}, {Poidevin}, {Liu}, {Coud{\'e}}, {Tahani}, {Liu}, {Onaka}, {Li}, {Tamura}, {Chen}, {Tang}, {Kirchschlager}, {Bourke}, {Byun}, {Chen}, {Chen}, {Chen}, {Cho}, {Choi}, {Choi}, {Choi}, {Chrysostomou}, {Chung}, {Dai}, {Diep}, {Doi}, {Duan}, {Duan}, {Eden}, {Fanciullo}, {Fiege}, {Fissel}, {Franzmann}, {Friberg}, {Friesen}, {Fuller}, {Gledhill}, {Graves}, {Greaves}, {Griffin}, {Gu}, {Han}, {Hayashi}, {Houde}, {Hull}, {Inoue}, {Inutsuka}, {Iwasaki}, {Jeong}, {K{\"o}nyves}, {Kang}, {Kang}, {Karoly}, {Kataoka}, {Kawabata}, {Kemper}, {Kim}, {Kim}, {Kim}, {Kim}, {Kim}, {Kim}, {Kirk}, {Kobayashi}, {Kusune}, {Kwon}, {Lacaille}, {Law}, {Lee}, {Lee}, {Lee}, {Lee}, {Lee}, {Li}, {Li}, {Lin}, {Liu}, {Lu}, {Mairs}, {Matsumura},
  {Matthews}, {Moriarty-Schieven}, {Nagata}, {Nakamura}, {Nakanishi}, {Ngoc}, {Ohashi}, {Park}, {Parsons}, {Peretto}, {Priestley}, {Pyo}, {Qian}, {Rao}, {Rawlings}, {Rawlings}, {Retter}, {Richer}, {Rigby}, {Sadavoy}, {Saito}, {Savini}, {Seta}, {Shimajiri}, {Shinnaga}, {Tang}, {Tomisaka}, {Tram}, {Tsukamoto}, {Viti}, {Wang}, {Wu}, {Xie}, {Yang}, {Yen}, {Yoo}, {Yuan}, {Yun}, {Zenko}, {Zhang}, {Zhang}, {Zhang}, {Zhou}, {Zhu}, {de Looze}, {Andr{\'e}}, {Dowell}, {Eyres}, {Falle}, {Robitaille}, \& {van Loo}}]{Chingetal2022}
{Ching}, T.-C., {Qiu}, K., {Li}, D., {et~al.} 2022, \apj, 941, 122, \dodoi{10.3847/1538-4357/ac9dfb}

\bibitem[{{Condon} {et~al.}(1998){Condon}, {Cotton}, {Greisen}, {Yin}, {Perley}, {Taylor}, \& {Broderick}}]{Condon1998}
{Condon}, J.~J., {Cotton}, W.~D., {Greisen}, E.~W., {et~al.} 1998, \aj, 115, 1693, \dodoi{10.1086/300337}

\bibitem[{{Crutcher}(1999)}]{Crutcher1999APJ520}
{Crutcher}, R.~M. 1999, \apj, 520, 706, \dodoi{10.1086/307483}

\bibitem[{{Crutcher}(2012)}]{Crutcher2012}
---. 2012, \araa, 50, 29, \dodoi{10.1146/annurev-astro-081811-125514}

\bibitem[{{Crutcher} \& {Kemball}(2019)}]{CrutcherKemball2019}
{Crutcher}, R.~M., \& {Kemball}, A.~J. 2019, Frontiers in Astronomy and Space Sciences, 6, 66, \dodoi{10.3389/fspas.2019.00066}

\bibitem[{{Crutcher} {et~al.}(1999{\natexlab{a}}){Crutcher}, {Roberts}, {Troland}, \& {Goss}}]{Crutcheretal1999APJ515}
{Crutcher}, R.~M., {Roberts}, D.~A., {Troland}, T.~H., \& {Goss}, W.~M. 1999{\natexlab{a}}, \apj, 515, 275, \dodoi{10.1086/307026}

\bibitem[{{Crutcher} {et~al.}(1996){Crutcher}, {Troland}, {Lazareff}, \& {Kazes}}]{Crutcheretal1996}
{Crutcher}, R.~M., {Troland}, T.~H., {Lazareff}, B., \& {Kazes}, I. 1996, \apj, 456, 217, \dodoi{10.1086/176642}

\bibitem[{{Crutcher} {et~al.}(1999{\natexlab{b}}){Crutcher}, {Troland}, {Lazareff}, {Paubert}, \& {Kaz{\`e}s}}]{Crutcher1999APJ514}
{Crutcher}, R.~M., {Troland}, T.~H., {Lazareff}, B., {Paubert}, G., \& {Kaz{\`e}s}, I. 1999{\natexlab{b}}, \apjl, 514, L121, \dodoi{10.1086/311952}

\bibitem[{{Crutcher} {et~al.}(2010){Crutcher}, {Wandelt}, {Heiles}, {Falgarone}, \& {Troland}}]{Crutcheretal2010}
{Crutcher}, R.~M., {Wandelt}, B., {Heiles}, C., {Falgarone}, E., \& {Troland}, T.~H. 2010, \apj, 725, 466, \dodoi{10.1088/0004-637X/725/1/466}

\bibitem[{{Doi} {et~al.}(2021){Doi}, {Hasegawa}, {Bastien}, {Tahani}, {Arzoumanian}, {Coud{\'e}}, {Matsumura}, {Sadavoy}, {Hull}, {Shimajiri}, {Furuya}, {Johnstone}, {Plume}, {Inutsuka}, {Kwon}, \& {Tamura}}]{Doietal2021}
{Doi}, Y., {Hasegawa}, T., {Bastien}, P., {et~al.} 2021, \apj, 914, 122, \dodoi{10.3847/1538-4357/abfcc5}

\bibitem[{{Doi} {et~al.}(2024){Doi}, {Nakamura}, {Kawabata}, {Matsumura}, {Akitaya}, {Coud{\'e}}, {Rodrigues}, {Kwon}, {Tamura}, {Tahani}, {Magalh{\~a}es}, {Santos-Lima}, {Angarita}, {Versteeg}, {Haverkorn}, {Hasegawa}, {Sadavoy}, {Arzoumanian}, \& {Bastien}}]{Doietal2024}
{Doi}, Y., {Nakamura}, K., {Kawabata}, K.~S., {et~al.} 2024, \apj, 961, 13, \dodoi{10.3847/1538-4357/ad0fe2}

\bibitem[{{Falgarone} {et~al.}(2008){Falgarone}, {Troland}, {Crutcher}, \& {Paubert}}]{Falgaroneetal2008}
{Falgarone}, E., {Troland}, T.~H., {Crutcher}, R.~M., \& {Paubert}, G. 2008, \aap, 487, 247, \dodoi{10.1051/0004-6361:200809577}

\bibitem[{{Gaensler} {et~al.}(2010){Gaensler}, {Landecker}, {Taylor}, \& {POSSUM Collaboration}}]{Gaensler2010}
{Gaensler}, B.~M., {Landecker}, T.~L., {Taylor}, A.~R., \& {POSSUM Collaboration}. 2010, in American Astronomical Society Meeting Abstracts, Vol. 215, American Astronomical Society Meeting Abstracts \#215, 470.13

\bibitem[{{Gibson} {et~al.}(2009){Gibson}, {Plume}, {Bergin}, {Ragan}, \& {Evans}}]{Gibsonetal2009}
{Gibson}, D., {Plume}, R., {Bergin}, E., {Ragan}, S., \& {Evans}, N. 2009, \apj, 705, 123, \dodoi{10.1088/0004-637X/705/1/123}

\bibitem[{Harris {et~al.}(2020)Harris, Millman, van~der Walt, Gommers, Virtanen, Cournapeau, Wieser, Taylor, Berg, Smith, Kern, Picus, Hoyer, van Kerkwijk, Brett, Haldane, del R{\'{i}}o, Wiebe, Peterson, G{\'{e}}rard-Marchant, Sheppard, Reddy, Weckesser, Abbasi, Gohlke, \& Oliphant}]{Numpy}
Harris, C.~R., Millman, K.~J., van~der Walt, S.~J., {et~al.} 2020, Nature, 585, 357, \dodoi{10.1038/s41586-020-2649-2}

\bibitem[{{Heald} {et~al.}(2020){Heald}, {Mao}, {Vacca}, {Akahori}, {Damas-Segovia}, {Gaensler}, {Hoeft}, {Agudo}, {Basu}, {Beck}, {Birkinshaw}, {Bonafede}, {Bourke}, {Bracco}, {Carretti}, {Feretti}, {Girart}, {Govoni}, {Green}, {Han}, {Haverkorn}, {Horellou}, {Johnston-Hollitt}, {Kothes}, {Landecker}, {Nikiel-Wroczy{\'n}ski}, {O'Sullivan}, {Padovani}, {Poidevin}, {Pratley}, {Regis}, {Riseley}, {Robishaw}, {Rudnick}, {Sobey}, {Stil}, {Sun}, {Sur}, {Taylor}, {Thomson}, {Van Eck}, {Vazza}, {West}, \& {SKA Magnetism Science Working Group}}]{Healdetal2020}
{Heald}, G., {Mao}, S., {Vacca}, V., {et~al.} 2020, Galaxies, 8, 53, \dodoi{10.3390/galaxies8030053}

\bibitem[{{Heiles}(1987)}]{Heiles1987}
{Heiles}, C. 1987, in Astrophysics and Space Science Library, Vol. 134, Interstellar Processes, ed. D.~J. {Hollenbach} \& H.~A. {Thronson}, Jr., 171--194, \dodoi{10.1007/978-94-009-3861-8_8}

\bibitem[{{Heiles}(1997)}]{Heiles1997}
{Heiles}, C. 1997, \apjs, 111, 245, \dodoi{10.1086/313010}

\bibitem[{{Heiles} \& {Robishaw}(2009)}]{PressRelease}
{Heiles}, C., \& {Robishaw}, T. 2009, in IAU Symposium, Vol. 259, Cosmic Magnetic Fields: From Planets, to Stars and Galaxies, ed. K.~G. {Strassmeier}, A.~G. {Kosovichev}, \& J.~E. {Beckman}, 579--590, \dodoi{10.1017/S174392130903141X}

\bibitem[{Hunter(2007)}]{Matplotlib}
Hunter, J.~D. 2007, Computing in Science \& Engineering, 9, 90, \dodoi{10.1109/MCSE.2007.55}

\bibitem[{{Hwang} {et~al.}(2022){Hwang}, {Kim}, {Pattle}, {Lee}, {Koch}, {Johnstone}, {Tomisaka}, {Whitworth}, {Furuya}, {Kang}, {Lyo}, {Chung}, {Arzoumanian}, {Park}, {Kwon}, {Kim}, {Tamura}, {Kwon}, {Soam}, {Han}, {Hoang}, {Kim}, {Onaka}, {Eswaraiah}, {Ward-Thompson}, {Liu}, {Tang}, {Chen}, {Matsumura}, {Hoang}, {Chen}, {Le Gouellec}, {Kirchschlager}, {Poidevin}, {Bastien}, {Qiu}, {Hasegawa}, {Lai}, {Byun}, {Cho}, {Choi}, {Choi}, {Choi}, {Jeong}, {Kang}, {Kim}, {Kim}, {Lee}, {Lee}, {Lee}, {Lee}, {Kim}, {Yoo}, {Yun}, {Chen}, {Di Francesco}, {Fiege}, {Fissel}, {Franzmann}, {Houde}, {Lacaille}, {Matthews}, {Sadavoy}, {Moriarty-Schieven}, {Tahani}, {Ching}, {Dai}, {Duan}, {Gu}, {Law}, {Li}, {Li}, {Li}, {Li}, {Liu}, {Lu}, {Qian}, {Wang}, {Wu}, {Xie}, {Yuan}, {Zhang}, {Zhang}, {Zhang}, {Zhou}, {Zhu}, {Berry}, {Friberg}, {Graves}, {Liu}, {Mairs}, {Parsons}, {Rawlings}, {Doi}, {Hayashi}, {Hull}, {Inoue}, {Inutsuka}, {Iwasaki}, {Kataoka}, {Kawabata}, {Kim}, {Kobayashi}, {Nagata}, {Nakamura}, {Nakanishi}, {Pyo},
  {Saito}, {Seta}, {Shimajiri}, {Shinnaga}, {Tsukamoto}, {Zenko}, {Chen}, {Duan}, {Fanciullo}, {Kemper}, {Lee}, {Lin}, {Liu}, {Ohashi}, {Rao}, {Tang}, {Wang}, {Yang}, {Yen}, {Bourke}, {Chrysostomou}, {Debattista}, {Eden}, {Eyres}, {Falle}, {Fuller}, {Gledhill}, {Greaves}, {Griffin}, {Hatchell}, {Karoly}, {Kirk}, {K{\"o}nyves}, {Longmore}, {van Loo}, {de Looze}, {Peretto}, {Priestley}, {Rawlings}, {Retter}, {Richer}, {Rigby}, {Savini}, {Scaife}, {Viti}, {Diep}, {Ngoc}, {Tram}, {Andr{\'e}}, {Coud{\'e}}, {Dowell}, {Friesen}, \& {Robitaille}}]{Hwangetal2022}
{Hwang}, J., {Kim}, J., {Pattle}, K., {et~al.} 2022, \apj, 941, 51, \dodoi{10.3847/1538-4357/ac99e0}

\bibitem[{{Kounkel} {et~al.}(2022){Kounkel}, {Deng}, \& {Stassun}}]{Kounkeletal2022}
{Kounkel}, M., {Deng}, T., \& {Stassun}, K.~G. 2022, arXiv e-prints, arXiv:2206.04703.
\newblock \doarXiv{2206.04703}

\bibitem[{{Krumholz} \& {Federrath}(2019)}]{KrumholzFederrath2019}
{Krumholz}, M.~R., \& {Federrath}, C. 2019, Frontiers in Astronomy and Space Sciences, 6, 7, \dodoi{10.3389/fspas.2019.00007}

\bibitem[{{Lazarian} \& {Hoang}(2007)}]{LazarianHoang2007}
{Lazarian}, A., \& {Hoang}, T. 2007, \mnras, 378, 910, \dodoi{10.1111/j.1365-2966.2007.11817.x}

\bibitem[{{Le Teuff} {et~al.}(2000){Le Teuff}, {Millar}, \& {Markwick}}]{LeTeuffetal1999}
{Le Teuff}, Y.~H., {Millar}, T.~J., \& {Markwick}, A.~J. 2000, \aaps, 146, 157, \dodoi{10.1051/aas:2000265}

\bibitem[{{O'Sullivan} {et~al.}(2023){O'Sullivan}, {Shimwell}, {Hardcastle}, {Tasse}, {Heald}, {Carretti}, {Br{\"u}ggen}, {Vacca}, {Sobey}, {Van Eck}, {Horellou}, {Beck}, {Bilicki}, {Bourke}, {Botteon}, {Croston}, {Drabent}, {Duncan}, {Heesen}, {Ideguchi}, {Kirwan}, {Lawlor}, {Mingo}, {Nikiel-Wroczy{\'n}ski}, {Piotrowska}, {Scaife}, \& {van Weeren}}]{O'Sullivanetal2023}
{O'Sullivan}, S.~P., {Shimwell}, T.~W., {Hardcastle}, M.~J., {et~al.} 2023, \mnras, 519, 5723, \dodoi{10.1093/mnras/stac3820}

\bibitem[{pandas~development team(2020)}]{Pandas}
pandas~development team, T. 2020, pandas-dev/pandas: Pandas, latest,  Zenodo, \dodoi{10.5281/zenodo.3509134}

\bibitem[{{Panopoulou} {et~al.}(2023){Panopoulou}, {Markopoulioti}, {Bouzelou}, {Millar-Blanchaer}, {Tinyanont}, {Blinov}, {Pelgrims}, {Johnson}, {Skalidis}, \& {Soam}}]{Panopoulouetal2023}
{Panopoulou}, G.~V., {Markopoulioti}, L., {Bouzelou}, F., {et~al.} 2023, arXiv e-prints, arXiv:2307.05752, \dodoi{10.48550/arXiv.2307.05752}

\bibitem[{{Pattle} \& {Fissel}(2019)}]{PattleFissel2019}
{Pattle}, K., \& {Fissel}, L. 2019, Frontiers in Astronomy and Space Sciences, 6, 15, \dodoi{10.3389/fspas.2019.00015}

\bibitem[{{Pattle} {et~al.}(2023){Pattle}, {Fissel}, {Tahani}, {Liu}, \& {Ntormousi}}]{Pattleetal2023PP7}
{Pattle}, K., {Fissel}, L., {Tahani}, M., {Liu}, T., \& {Ntormousi}, E. 2023, in Astronomical Society of the Pacific Conference Series, Vol. 534, Protostars and Planets VII, ed. S.~{Inutsuka}, Y.~{Aikawa}, T.~{Muto}, K.~{Tomida}, \& M.~{Tamura}, 193, \dodoi{10.48550/arXiv.2203.11179}

\bibitem[{Pedregosa {et~al.}(2011)Pedregosa, Varoquaux, Gramfort, Michel, Thirion, Grisel, Blondel, Prettenhofer, Weiss, Dubourg, {et~al.}}]{sklearn}
Pedregosa, F., Varoquaux, G., Gramfort, A., {et~al.} 2011, Journal of machine learning research, 12, 2825

\bibitem[{{Planck Collaboration} {et~al.}(2015){Planck Collaboration}, {Ade}, {Aghanim}, {Alina}, {Alves}, {Armitage-Caplan}, {Arnaud}, {Arzoumanian}, {Ashdown}, {Atrio-Barandela}, {Aumont}, {Baccigalupi}, {Banday}, {Barreiro}, {Battaner}, {Benabed}, {Benoit-L{\'e}vy}, {Bernard}, {Bersanelli}, {Bielewicz}, {Bock}, {Bond}, {Borrill}, {Bouchet}, {Boulanger}, {Bracco}, {Burigana}, {Butler}, {Cardoso}, {Catalano}, {Chamballu}, {Chary}, {Chiang}, {Christensen}, {Colombi}, {Colombo}, {Combet}, {Couchot}, {Coulais}, {Crill}, {Curto}, {Cuttaia}, {Danese}, {Davies}, {Davis}, {de Bernardis}, {de Gouveia Dal Pino}, {de Rosa}, {de Zotti}, {Delabrouille}, {D{\'e}sert}, {Dickinson}, {Diego}, {Donzelli}, {Dor{\'e}}, {Douspis}, {Dunkley}, {Dupac}, {Efstathiou}, {En{\ss}lin}, {Eriksen}, {Falgarone}, {Ferri{\`e}re}, {Finelli}, {Forni}, {Frailis}, {Fraisse}, {Franceschi}, {Galeotta}, {Ganga}, {Ghosh}, {Giard}, {Giraud-H{\'e}raud}, {Gonz{\'a}lez-Nuevo}, {G{\'o}rski}, {Gregorio}, {Gruppuso}, {Guillet}, {Hansen}, {Harrison},
  {Helou}, {Hern{\'a}ndez-Monteagudo}, {Hildebrandt}, {Hivon}, {Hobson}, {Holmes}, {Hornstrup}, {Huffenberger}, {Jaffe}, {Jaffe}, {Jones}, {Juvela}, {Keih{\"a}nen}, {Keskitalo}, {Kisner}, {Kneissl}, {Knoche}, {Kunz}, {Kurki-Suonio}, {Lagache}, {L{\"a}hteenm{\"a}ki}, {Lamarre}, {Lasenby}, {Lawrence}, {Leahy}, {Leonardi}, {Levrier}, {Liguori}, {Lilje}, {Linden-V{\o}rnle}, {L{\'o}pez-Caniego}, {Lubin}, {Mac{\'\i}as-P{\'e}rez}, {Maffei}, {Magalh{\~a}es}, {Maino}, {Mandolesi}, {Maris}, {Marshall}, {Martin}, {Mart{\'\i}nez-Gonz{\'a}lez}, {Masi}, {Matarrese}, {Mazzotta}, {Melchiorri}, {Mendes}, {Mennella}, {Migliaccio}, {Miville-Desch{\^e}nes}, {Moneti}, {Montier}, {Morgante}, {Mortlock}, {Munshi}, {Murphy}, {Naselsky}, {Nati}, {Natoli}, {Netterfield}, {Noviello}, {Novikov}, {Novikov}, {Oxborrow}, {Pagano}, {Pajot}, {Paladini}, {Paoletti}, {Pasian}, {Pearson}, {Perdereau}, {Perotto}, {Perrotta}, {Piacentini}, {Piat}, {Pietrobon}, {Plaszczynski}, {Poidevin}, {Pointecouteau}, {Polenta}, {Popa}, {Pratt}, {Prunet},
  {Puget}, {Rachen}, {Reach}, {Rebolo}, {Reinecke}, {Remazeilles}, {Renault}, {Ricciardi}, {Riller}, {Ristorcelli}, {Rocha}, {Rosset}, {Roudier}, {Rubi{\~n}o-Mart{\'\i}n}, {Rusholme}, {Sandri}, {Savini}, {Scott}, {Spencer}, {Stolyarov}, {Stompor}, {Sudiwala}, {Sutton}, {Suur-Uski}, {Sygnet}, {Tauber}, {Terenzi}, {Toffolatti}, {Tomasi}, {Tristram}, {Tucci}, {Umana}, {Valenziano}, {Valiviita}, {Van Tent}, {Vielva}, {Villa}, {Wade}, {Wandelt}, {Zacchei}, \& {Zonca}}]{PlanckXIX2015}
{Planck Collaboration}, {Ade}, P.~A.~R., {Aghanim}, N., {et~al.} 2015, \aap, 576, A104, \dodoi{10.1051/0004-6361/201424082}

\bibitem[{{Planck Collaboration} {et~al.}(2016){Planck Collaboration}, {Ade}, {Aghanim}, {Alves}, {Arnaud}, {Arzoumanian}, {Ashdown}, {Aumont}, {Baccigalupi}, {Banday}, {Barreiro}, {Bartolo}, {Battaner}, {Benabed}, {Beno{\^i}t}, {Benoit-L{\'e}vy}, {Bernard}, {Bersanelli}, {Bielewicz}, {Bock}, {Bonavera}, {Bond}, {Borrill}, {Bouchet}, {Boulanger}, {Bracco}, {Burigana}, {Calabrese}, {Cardoso}, {Catalano}, {Chiang}, {Christensen}, {Colombo}, {Combet}, {Couchot}, {Crill}, {Curto}, {Cuttaia}, {Danese}, {Davies}, {Davis}, {de Bernardis}, {de Rosa}, {de Zotti}, {Delabrouille}, {Dickinson}, {Diego}, {Dole}, {Donzelli}, {Dor{\'e}}, {Douspis}, {Ducout}, {Dupac}, {Efstathiou}, {Elsner}, {En{\ss}lin}, {Eriksen}, {Falceta-Gon{\c c}alves}, {Falgarone}, {Ferri{\`e}re}, {Finelli}, {Forni}, {Frailis}, {Fraisse}, {Franceschi}, {Frejsel}, {Galeotta}, {Galli}, {Ganga}, {Ghosh}, {Giard}, {Gjerl{\o}w}, {Gonz{\'a}lez-Nuevo}, {G{\'o}rski}, {Gregorio}, {Gruppuso}, {Gudmundsson}, {Guillet}, {Harrison}, {Helou}, {Hennebelle},
  {Henrot-Versill{\'e}}, {Hern{\'a}ndez-Monteagudo}, {Herranz}, {Hildebrandt}, {Hivon}, {Holmes}, {Hornstrup}, {Huffenberger}, {Hurier}, {Jaffe}, {Jaffe}, {Jones}, {Juvela}, {Keih{\"a}nen}, {Keskitalo}, {Kisner}, {Knoche}, {Kunz}, {Kurki-Suonio}, {Lagache}, {Lamarre}, {Lasenby}, {Lattanzi}, {Lawrence}, {Leonardi}, {Levrier}, {Liguori}, {Lilje}, {Linden-V{\o}rnle}, {L{\'o}pez-Caniego}, {Lubin}, {Mac{\'{\i}}as-P{\'e}rez}, {Maino}, {Mandolesi}, {Mangilli}, {Maris}, {Martin}, {Mart{\'{\i}}nez-Gonz{\'a}lez}, {Masi}, {Matarrese}, {Melchiorri}, {Mendes}, {Mennella}, {Migliaccio}, {Miville-Desch{\^e}nes}, {Moneti}, {Montier}, {Morgante}, {Mortlock}, {Munshi}, {Murphy}, {Naselsky}, {Nati}, {Netterfield}, {Noviello}, {Novikov}, {Novikov}, {Oppermann}, {Oxborrow}, {Pagano}, {Pajot}, {Paladini}, {Paoletti}, {Pasian}, {Perotto}, {Pettorino}, {Piacentini}, {Piat}, {Pierpaoli}, {Pietrobon}, {Plaszczynski}, {Pointecouteau}, {Polenta}, {Ponthieu}, {Pratt}, {Prunet}, {Puget}, {Rachen}, {Reinecke}, {Remazeilles}, {Renault},
  {Renzi}, {Ristorcelli}, {Rocha}, {Rossetti}, {Roudier}, {Rubi{\~n}o-Mart{\'{\i}}n}, {Rusholme}, {Sandri}, {Santos}, {Savelainen}, {Savini}, {Scott}, {Soler}, {Stolyarov}, {Sudiwala}, {Sutton}, {Suur-Uski}, {Sygnet}, {Tauber}, {Terenzi}, {Toffolatti}, {Tomasi}, {Tristram}, {Tucci}, {Umana}, {Valenziano}, {Valiviita}, {Van Tent}, {Vielva}, {Villa}, {Wade}, {Wandelt}, {Wehus}, {Ysard}, {Yvon}, \& {Zonca}}]{PlanckXXXV}
---. 2016, \aap, 586, A138, \dodoi{10.1051/0004-6361/201525896}

\bibitem[{{Pudritz} \& {Ray}(2019)}]{PudritzRay2019}
{Pudritz}, R.~E., \& {Ray}, T.~P. 2019, Frontiers in Astronomy and Space Sciences, 6, 54, \dodoi{10.3389/fspas.2019.00054}

\bibitem[{{Robishaw}(2008)}]{RobishawThesis}
{Robishaw}, T. 2008, PhD thesis, University of California, Berkeley

\bibitem[{{Seifried} \& {Walch}(2015)}]{SeifriedWalch2015}
{Seifried}, D., \& {Walch}, S. 2015, \mnras, 452, 2410, \dodoi{10.1093/mnras/stv1458}

\bibitem[{{Seifried} {et~al.}(2019){Seifried}, {Walch}, {Reissl}, \& {Ib{\'a}{\~n}ez-Mej{\'\i}a}}]{Seifriedetal2019}
{Seifried}, D., {Walch}, S., {Reissl}, S., \& {Ib{\'a}{\~n}ez-Mej{\'\i}a}, J.~C. 2019, \mnras, 482, 2697, \dodoi{10.1093/mnras/sty2831}

\bibitem[{{Tahani}(2022)}]{Tahani2022}
{Tahani}, M. 2022, Frontiers in Astronomy and Space Sciences, 9, 940027, \dodoi{10.3389/fspas.2022.940027}

\bibitem[{{Tahani} {et~al.}(2018){Tahani}, {Plume}, {Brown}, \& {Kainulainen}}]{Tahanietal2018}
{Tahani}, M., {Plume}, R., {Brown}, J.~C., \& {Kainulainen}, J. 2018, \aap, 614, A100, \dodoi{10.1051/0004-6361/201732219}

\bibitem[{{Tahani} {et~al.}(2019){Tahani}, {Plume}, {Brown}, {Soler}, \& {Kainulainen}}]{Tahanietal2019}
{Tahani}, M., {Plume}, R., {Brown}, J.~C., {Soler}, J.~D., \& {Kainulainen}, J. 2019, \aap, 632, A68, \dodoi{10.1051/0004-6361/201936280}

\bibitem[{{Tahani} {et~al.}(2022{\natexlab{a}}){Tahani}, {Glover}, {Lupypciw}, {West}, {Kothes}, {Plume}, {Inutsuka}, {Lee}, {Grenier}, {Knee}, {Brown}, {Doi}, {Robishaw}, \& {Haverkorn}}]{Tahanietal2022O}
{Tahani}, M., {Glover}, J., {Lupypciw}, W., {et~al.} 2022{\natexlab{a}}, \aap, 660, L7, \dodoi{10.1051/0004-6361/202243322}

\bibitem[{{Tahani} {et~al.}(2022{\natexlab{b}}){Tahani}, {Lupypciw}, {Glover}, {Plume}, {West}, {Kothes}, {Inutsuka}, {Lee}, {Robishaw}, {Knee}, {Brown}, {Doi}, {Grenier}, \& {Haverkorn}}]{Tahanietal2022P}
{Tahani}, M., {Lupypciw}, W., {Glover}, J., {et~al.} 2022{\natexlab{b}}, \aap, 660, A97, \dodoi{10.1051/0004-6361/202141170}

\bibitem[{{Tahani} {et~al.}(2023){Tahani}, {Bastien}, {Furuya}, {Pattle}, {Johnstone}, {Arzoumanian}, {Doi}, {Hasegawa}, {Inutsuka}, {Coud{\'e}}, {Fissel}, {Chen}, {Poidevin}, {Sadavoy}, {Friesen}, {Koch}, {Di Francesco}, {Moriarty-Schieven}, {Chen}, {Chung}, {Eswaraiah}, {Fanciullo}, {Gledhill}, {Le Gouellec}, {Hoang}, {Hwang}, {Kang}, {Kim}, {Kirchschlager}, {Kwon}, {Lee}, {Liu}, {Onaka}, {Rawlings}, {Soam}, {Tamura}, {Tang}, {Tomisaka}, {Whitworth}, {Kwon}, {Hoang}, {Redman}, {Berry}, {Ching}, {Wang}, {Lai}, {Qiu}, {Ward-Thompson}, {Houde}, {Byun}, {Chen}, {Chen}, {Cho}, {Choi}, {Choi}, {Chrysostomou}, {Diep}, {Duan}, {Fiege}, {Franzmann}, {Friberg}, {Fuller}, {Graves}, {Greaves}, {Griffin}, {Gu}, {Han}, {Hatchell}, {Hayashi}, {Hull}, {Inoue}, {Iwasaki}, {Jeong}, {Kanamori}, {Kang}, {Kang}, {Kataoka}, {Kawabata}, {Kemper}, {Kim}, {Kim}, {Kim}, {Kim}, {Kim}, {Kirk}, {Kobayashi}, {Konyves}, {Kusune}, {Lacaille}, {Law}, {Lee}, {Lee}, {Lee}, {Lee}, {Lee}, {Li}, {Li}, {Li}, {Liu}, {Liu}, {Liu}, {de Looze},
  {Lyo}, {Mairs}, {Matsumura}, {Matthews}, {Nagata}, {Nakamura}, {Nakanishi}, {Ohashi}, {Park}, {Parsons}, {Peretto}, {Pyo}, {Qian}, {Rao}, {Retter}, {Richer}, {Rigby}, {Saito}, {Savini}, {Scaife}, {Seta}, {Shimajiri}, {Shinnaga}, {Tang}, {Tsukamoto}, {Viti}, {Wang}, {Yen}, {Yoo}, {Yuan}, {Yun}, {Zenko}, {Zhang}, {Zhang}, {Zhang}, {Zhou}, {Zhu}, {Andr{\'e}}, {Dowell}, {Eyres}, {Falle}, {van Loo}, \& {Robitaille}}]{Tahanietal2023}
{Tahani}, M., {Bastien}, P., {Furuya}, R.~S., {et~al.} 2023, \apj, 944, 139, \dodoi{10.3847/1538-4357/acac81}

\bibitem[{{Taylor} {et~al.}(2009){Taylor}, {Stil}, \& {Sunstrum}}]{Tayloretal2009}
{Taylor}, A.~R., {Stil}, J.~M., \& {Sunstrum}, C. 2009, \apj, 702, 1230, \dodoi{10.1088/0004-637X/702/2/1230}

\bibitem[{{Van Eck} {et~al.}(2017){Van Eck}, {Haverkorn}, {Alves}, {Beck}, {de Bruyn}, {En{\ss}lin}, {Farnes}, {Ferri{\`e}re}, {Heald}, {Horellou}, {Horneffer}, {Iacobelli}, {Jeli{\'c}}, {Mart{\'{\i}}-Vidal}, {Mulcahy}, {Reich}, {R{\"o}ttgering}, {Scaife}, {Schnitzeler}, {Sobey}, \& {Sridhar}}]{VanEcketal2017}
{Van Eck}, C.~L., {Haverkorn}, M., {Alves}, M.~I.~R., {et~al.} 2017, \aap, 597, A98, \dodoi{10.1051/0004-6361/201629707}

\bibitem[{{Van Eck} {et~al.}(2023){Van Eck}, {Gaensler}, {Hutschenreuter}, {Livingston}, {Ma}, {Riseley}, {Thomson}, {Adebahr}, {Basu}, {Birkinshaw}, {En{\ss}lin}, {Heald}, {Mao}, \& {McClure-Griffiths}}]{VanEcketal2023}
{Van Eck}, C.~L., {Gaensler}, B.~M., {Hutschenreuter}, S., {et~al.} 2023, \apjs, 267, 28, \dodoi{10.3847/1538-4365/acda24}

\bibitem[{Van~Rossum(2020)}]{Mathpython}
Van~Rossum, G. 2020, The Python Library Reference, release 3.8.2 (Python Software Foundation)

\bibitem[{{Vanderwoude} {et~al.}(2024){Vanderwoude}, {West}, {Gaensler}, {Rudnick}, {Van Eck}, {Thomson}, {Andernach}, {Anderson}, {Carretti}, {Heald}, {Leahy}, {McClure-Griffiths}, {O'Sullivan}, {Tahani}, \& {Willis}}]{Vanderwoudeetal2024}
{Vanderwoude}, S., {West}, J.~L., {Gaensler}, B.~M., {et~al.} 2024, \aj, 167, 226, \dodoi{10.3847/1538-3881/ad2fc8}

\bibitem[{{Wolleben} \& {Reich}(2004)}]{WollebenReich2004Molecular}
{Wolleben}, M., \& {Reich}, W. 2004, \aap, 427, 537, \dodoi{10.1051/0004-6361:20040561}

\end{thebibliography}
\bibliographystyle{aasjournal}

\end{document}